# GDP growth rate and population

Ivan O. Kitov

## Abstract

Real GDP growth rate in developed countries is found to be a sum of two terms. The first term is the reciprocal value of the duration of the period of mean income growth with work experience, $T_{cr}$. The current value of $T_{cr}$ in the USA is 40 years. The second term is inherently related to population and defined by the relative change in the number of people with a specific age (9 years in the USA),

*(1/2)*dN9(t) /N9(t),*

where N9(t) is the number of 9-year-olds at time t. The $T_{cr}$ grows as the square root of real GDP per capita.

Hence, evolution of real GDP is defined by only one parameter - the number of people of the specific age. Predictions for the USA, the UK, and France are presented and discussed.

A similar relationship is derived for real GDP per capita. Annual increment of GDP per capita is also a combination of economic trend term and the same specific age population term. The economic trend term during last 55 years is equal to $400 (2002 US dollars) divided by the attained level of real GDP per capita. Thus, the economic trend term has an asymptotic value of zero.

Inversion of the measured GDP values is used to recover the corresponding change of the specific age population between 1955 and 2003. The population recovery method based on GDP potentially is of a higher accuracy than routine censuses.

## Introduction

A comprehensive study of the US personal income distribution (PID) and detailed modelling of some important characteristics of the distribution is carried out by Kitov (2005a). The principal finding is that people as economic agents producing (equivalent - earning) money are distributed according to a fixed and hierarchical structure resulting in a very rigid response of the personal income distribution to any external disturbances including inflation and real economic growth. There is a predefined distribution of relative income, i.e. portion of the total population obtaining a given portion of the total real income. In addition, every place in the distribution is occupied by somebody. A person occupying a given place may propagate to a position with a different income, but the vacant place must be filled by somebody. For example, by the person who was in the new place of the first one. Only such an exchange of income positions in the PID, or more complicated change of positions with circular substitutes, is possible. This mechanism provides a dynamic equilibrium and the observed stable personal income distribution.

The measured PIDs in the USA corrected for the observed nominal per capita GDP growth rate show a very stable shape during the period between 1994 and 2003. This stability is interpreted as an existence of an almost stable relative income distribution hierarchy in American society, which might be developing very slowly with time. Then, inflation should represent a mechanism compensating disturbance of the PID caused by real economic growth. Inflation eats out of the poor people advantages obtained from the real economic growth.

The economic structure also predefines the observed economic evolution. Only characteristics of age distribution in the population are important for GDP growth rate above some economic trend. Numerically, the latter is inversely proportional to the attained value of per capita real GDP. Analysis of the two factors of the American economic growth is the main goal of this paper. The analysis is focused on the decomposition of real economic growth (GDP) and per capita real economic growth in developed economic countries into an economic trend and fluctuations as described by theories of business cycles proposed by Hodrick and Prescott (1980). The sense of the two terms is different, however.

## 1. The model for the prediction of GDP growth rate

Per capita GDP growth rate in the USA was used by Kitov (2005a) as an external parameter in prediction of the observed evolution of the PID, its components and derivatives. The PID has been expressed as a simple and predetermined function of GDP per capita and the age structure of the working age population in the USA. The current study interprets this relationship in the reverse direction. The observed PID is considered as a result of each and every individual effort to earn (equivalent - to produce goods and services) money in the economically structured society as exists in the USA. Thus, the individual money production (earning) aggregated over the US working age population is the inherent driving force of the observed economic development. The working age means the age eligible to receive income, i.e. 15 years of age and above. This effectively includes all retired people.

The principal assumption made by Kitov (2005a) and retained in the current study is that GDP denominated in money is the sum of all the personal incomes of all the people over 14 years of age. This statement not only formulates the income side of GDP definition but extends Walrasian equilibrium to all people above 14 years of age, with income being the only measure of the produced goods and services whatever they are. This statement unambiguously defines the upper limit to the total income (Gross Domestic Income) or GDP which can be produced by a population with a given age structure and characterized by some attained level of GDP per capita. As the age structure is given and individual incomes in the society are predefined by a strict relationship between age and per capita GDP (Kitov, 2005a), the total potential income growth has to be also predefined.

By definition, a person produces exactly the same amount of money (as goods, services, or something else) as s/he receives as income. This provides a global balance of income (earnings) and production, but also a more strict and important local balance. Economic structure of a developed economic society confines its possible evolution as everybody has an income place (position in PID) and produces according to this place.

This approach also implies that there is no economic means to disturb the economic structure of such a society. For example, it is impossible to reduce poverty or to limit individual incomes of rich people compared to the level predefined by the economic structure itself. According to the PID observations in the USA, all positions of poor and rich people in the structure are always occupied. This might be not the case in other countries. The extent to which the positions are occupied can be potentially linked to the degree of economic performance. Performance in a disturbed income structure should be reduced compared to its potential level only defined by per capita GDP and age structure. Thus, only some non-economic means are available to fight poverty. A society can provide higher living standards but not higher incomes if it does not wand to lose economic competitiveness. When applied, any economic means (income redistribution in favor of poorer people) have to result in economic underperformance. Another possibility is that some mechanisms out of control will return the PID to its original shape with the same number of poor people.

Per capita GDP growth rate is uniquely determined by the current distribution of the personal income which, in turn, depends on population age distribution. As shown by Kitov (2005b), the mean personal income distribution is only governed by two values – at the starting point of the distribution and the $T_{cr}$ - the value of work experience characterized by the highest mean income. Integral of the product of the mean personal income and the number of people with given work experience over the work experience range gives a GDP estimate. Thus, one can assume that the numerical value of real GDP growth rate in developed countries, which are characterized by a stable economic structure or PID, can be represented as a sum of two terms. The first term is the reciprocal value of the $T_{cr}$, which is often called economic trend or potential. Current value (2004) of $T_{cr}$ in the USA is 40 years, i.e. the current economic trend, including the working age population growth, is 0.025. The second term is inherently related to the number of young people of some specific age. This defining age has to be determined by calibration and may vary with country. In the USA and the UK, the age is nine years. In European countries and Japan it reaches eighteen years. This population related term creates the observed high frequency fluctuations in GDP growth rate and is expressed by the following relationship: $0.5*dN(t))/N(t)dt,$ where $N(t)$ is the number of people with the specific age at time $t$. Thus, one can write the following relationship for the GDP change:

$$g(t)=dG(t)/G(t)dt=0.5dN(t)/N(t)dt+1/T_{cr}(t) \quad \textbf{(1)}$$

where $g(t)$ is the real GDP growth rate, G(t) is the real GDP as a function of time.

Completing the system of equations is the relationship between the growth rate of per capita real GDP and the $T_{cr}$:

$$T_{cr}(t)=T_{cr}(t_0)\ sqrt(\int(1+g(t)-n(t))dt) \quad \textbf{(2)}$$

where $n(t)=dNT(t)/NT(t)dt$ is the time derivative of the total working age population relative change during the same period of time. The term in brackets under integral is the per capita real GDP growth rate. So, the critical work experience evolves in time as the square root of the total real per capita GDP gain between the starting time $t_0$ and $t$.

## 2. GDP growth rate prediction

Using equations (1) and (2) one can predict the observed evolution of real GDP in the USA. Systematic substitution of various single year of age time series in (1) gives the best fit for nine years of age. The single year population estimates used in the study are available at the U.S. Census Bureau web-site (2004a–c). There are several different sets of population data which undergo revisions as new information or methodology becomes available. Censuses are carried out every ten years and the last one was conducted in 2000. After the 2000 census, all the population estimates for the period from 1990 to 2000 were adjusted for matching the new census counts (US Census Bureau, 2005d). The difference between the estimated and counted population at April 1, 2000 is called “the closure error”. The population estimates made for this period were based on results of the 1990 census and the measured change in population components during this decade. These population estimates showed sometimes very poor results compared to the counting. Thus, there are two data sets for the period: an intercensal set and a postcensal one, as defined by the U.S. Census Bureau. For the period before 1990 only one data set is available. Supposedly, this is an intercensal estimate that uses data from corresponding bounding censuses.

Figures 1a and 1b present the measured (BEA, 2005) and predicted by (1) and (2) real GDP growth rate values. The number of 9-year-olds is taken from the intercensal and postcensal estimates respectively (US CB, 2005d). Overall, the measured and derived time series are not in a good agreement. A relatively good agreement is observed only near 1980 and 2000.

The used intercensal single year of age population estimate is adjusted to the results of decennial censuses. So, the time series has to provide the best estimates. There are some indications, however, that the data are over-smoothed and corrected in a way to suppress any possible large change in the single year population. As stated in the overview of the population estimates (US CB, 2005e) - "These [population] estimates are used in federal funding allocations, as denominators for vital rates and per capita time series, as survey controls, and in monitoring recent demographic changes. ". This goal is slightly different from that for accurate prediction of the number of nine year old children needed for the GDP prediction.

One can expect that the population distributions obtained in censuses are not biased by corrections, adjustments and modifications to the extent the post- and intercensal estimates are. A single year of age distribution can be obtained from an age pyramid as a projection back and forth in time. For example, the number of nine-year-olds in the next year from a census year is considered as equal to the number of eight-years-olds in the census year, and so on. The larger time gap between the census and the predicted year the larger is the error induced by all the demographic changes during these years. The same projection procedure can be applied to each of the available annual distributions.

Because the model uses a relative rather than absolute population change in adjacent years, the bias introduced into the projection procedure by demographic processes might be not so large. For example, one can consider a process of 1% population growth per year. This value is close to the total population change observed in the USA during the last 40 years. If every single year of age population grows at the same rate, the ratio of the adjacent single year of age populations is not affected by the population growth. If the adjacent single-year populations $a(n)$ and $a(n+1)$ (of age *n* and *n+1*) undergo a constant absolute growth by *p* persons every year, then the corresponding ratio will be of $a(n+m)/a(n+1+m) \approx 1+r(1-mp/a(n))$ in m years, where *r* is the initial ratio. If the absolute growth mp<<a then the approximation gives a relatively high accuracy. If the absolute change is not synchronized between adjacent single years of age cohorts, the ratio may be disturbed.

Thus, one can expect a much smaller relative change between adjacent cohorts than the absolute one. The degree of the relative change is not well determined, however, because of a relatively low accuracy of the measurements used in the population estimates. The accuracy may be evaluated by a comparison of the postcensal estimates and results of the decennial censuses. The difference between the predicted from continuous observations and counted numbers is as high as several percent and is particularly high for the ages between 5 and 10 (West and Robinson, 1997).

The estimates of the United States population are currently derived at a quarterly rate by updating the resident population enumerated in censuses through the components of the population change. The following components are used: census enumeration of resident population+ births to U.S. resident women - deaths to U.S. residents + net international migration + net movement of U.S. Armed Forces. This estimate is carried out for every single year of age, race and gender. The former is of importance for the study.

Figures 2 through 5 compare the readings of real GDP growth rate and those predicted from the 9-year-old population estimates based on censuses – from 1980 to 2000. Figure 2 illustrates how well one can predict the real GDP growth rate around 1980. The predicted and observed curves are very close and have similar shape, but of slightly different amplitude. The curve predicted from the 1980 population estimate (the census year estimate) is much closer to the real GDP growth rate curve than the curved obtained from the 9-year-olds estimates as presented in Figure 1. This discrepancy is consistent between the curves based on particular censuses and on the 9-year-olds estimates obtained with later corrections and modifications.

Figure 3 demonstrates predictive power of the 1990 census. The 9-year-olds estimate (see Figure 1) for the period around 1990 shows very poor correlation between the observed and predicted GDP growth rate. One can conclude that the curves can not represent the same process due to a substantial difference in amplitude and shape. This is quite opposite to the prediction based on the 1990 population enumeration. Amplitude and shape of the predicted GDP curve are almost the same as those observed near 1990. The most important observation is that the census population estimate unambiguously indicates the observed recession in 1991. Moreover, the drop could be predicted several years before if the population estimates between 1982 and 1991 would be accurate enough. Such a prediction is of the highest interest for business because it allows forecast of inevitable recessions.

The same is valid for the 2000 population estimates. The Census Bureau gives two estimates for the period between 1990 and 2000: postcensal and intercensal. Figures 4 and 5 display the GDP predictions based on the intercensal and postcensal population estimates respectively. The intercensal estimate gives a good timing of the 2001 recession, but underestimates amplitude of the drop. The postcensal estimate gives the same timing but slightly overestimates the amplitude. The estimates also give a very good timing and relatively accurate amplitude for the two previous recession periods. Thus, at least the three previous recession periods are well described by the 9-year-old population change as described by relationship (1) if to use the enumerated age distributions instead of those modified by the Census Bureau.

The intercensal estimate represents a corrected version of the postcensal one and considers not only continuous data on birth rate, mortality rate and international migration, but also the final distribution obtained in the 2000 enumeration. One can find a typical trend in the intercensal estimate which smoothing out the population changes between neighbouring years of age. Figure 6 illustrates this finding by a comparison of the two predicted GDP time series. There is no actual population age distribution available for the estimation of the absolute accuracy of the postcensal or intercensal estimates, i.e. the difference between the actual and enumerated numbers. One can only estimate some relative changes made by the Census Bureau and compare those to the observed discrepancy between the real GDP growth rate estimate and the predicted from the 9-year-olds estimates.

One should also bear in mind that the GDP estimates are also exposed to revisions induced by late data arrivals and methodology changes. The corrections may be as high as 1 percentage point and be extended by 10 years back in past. For example, the US Bureau of Economic Analysis published in July 2005 some corrections to the GDP estimates for the previous years that considerably changed the GDP values between 2001 and 2004. Presumably, the officially published GDP estimates are only 0.5 to 1 percentage point accurate and any discrepancy of such an amplitude between the observed and predicted GDP values might be induced by the real GDP estimate uncertainty and not only the population miscount.

## 3. Accuracy of the population estimates

Accuracy of the population estimates used in the GDP prediction is of a crucial importance for the study. The prediction of the recession periods between 1980 and 2004 in time and amplitude is possible only due to the fact that the observed changes were large enough and can not be smoothed out even by the Census Bureau adjustments. Between the census years, however, one would like to have a better agreement between the observations and predictions.

There are two main questions in the study related to the accuracy of the population estimates:

- Is it possible to recover the actual number of 9-year-olds from the Census Bureau population estimates?

If not,

- Does the 9-year-olds population, which accurately predicts the observed real GDP growth rate, contradict the Census Bureau population estimates?

There are two general sources of error in the population estimates. The first is the census miscount induced by many reasons. The postcensal estimates of accuracy conducted by the US Census Bureau undoubtedly show that the accuracy of the enumeration depends on age. The largest miscount corresponds usually to the age range between 5 and 10 years. The highest relative accuracy of the enumeration is observed in the age range between 20 and 30 years. The miscount of children is an underestimation as a rule. It can reach 5 percentage points (West and Robinson, 1997).

Figures 7 and 8 demonstrate the changes made by the Census Bureau to several single year of age populations. Figure 7 displays evolution in time of the postcensal estimates for 7-, 8-, 9-, and 10-year-olds. The curves are shifted relative to that corresponding to the 9-year-olds by a number of years equal to the difference of the ages. For example the curve for 7-year-olds is shifted by 2=9-7 years forward in order to trace the same single year of age cohort. The order on the curves looks suspicious. The number of 8-year-olds is by almost 300,000 less than the number of 9-year-olds in the next year and 150,000 less than the number of 7-year-olds in the previous year. This means that the number of people in this age cohort initially drops by 150,000 (from 7 to 8 years of age) and then increases by 300,000 in the next year. For improvement of this strange behavior, the Census Bureau has made considerable intercensal correction of the numbers between 1990 and 2000 as shown in Figure 8. The difference between 7-year-olds and 8-year-olds is gradually decreased from 150,000 in 1990 to almost a zero value in 2000, and then the 8-year-olds curve goes above the 7-year-olds one. Both curves are approaching the 9-year-olds curve between 1990 and 2000. The procedure for the correction implied proportional decrease of the population difference, i.e. a constant rate of catch up. This implies that no possible changes induced by the changes in birth and mortality rate, net international migration are taken into account in the correction process. One can conclude that there is no possibility to recover the actual 9-year-olds time series from the US Census Bureau data.

## 4. Population estimates by the real GDP evolution

One can easily find the 9-year-old population corresponding to the observed GDP growth rate. The number is described by the following relationship which represents an inversed version of (1):

$$d( lnN(t)) =2(g-1/Tcr))dt \quad \textbf{(3)}$$

The number of 9-year-olds in a given year *i* is equal to the number in the previous year *i*-1 times the term, which reflects the difference between the observed real GDP growth rate, $g(t)$, and the potential growth rate corresponding to the observed real GDP per capita value as defined by relationship (2). This working experience or the critical time, $T_{cr}$, depends only on the real GDP per capita. Thus, the term in brackets is completely defined by the real GDP measurements. It is worth noting that the $T_{cr}$ can be determined independently from the personal income distribution reports of the Census Bureau (Kitov, 2005b). The term in brackets is very sensitive to the evolution of $T_{cr}$ in time. If the observed long-term behaviour of GDP is not accurately described by the term $1/T_{cr}$, discrepancy between the observed and predicted number of 9-year-olds has to be large.

The only value defining the $N(t)$, considering $g(t)$ is obtained from an independent set of measurements, is its initial value for some given year, $N(t_0)$. One can vary $N(t_0)$ in order to reach the best fit between the Census Bureau 9-year-olds population estimates and the value predicted by relationship (3). In the study, some year between 1950 and 1960 is selected. The principal goal of the fitting process was to reach the best agreement during the last 10 to 30 years. The period before 1970 is characterized by a lower accuracy of the population estimates whether made by the Census Bureau or back projected from the population age pyramid for some specific years.

Figures 11 through 16 present a series of comparisons between the estimated and predicted number of 9-year-olds for the years between 1960 and 2001. The bounds of the population estimate uncertainty adopted in the study is 5%. Every comparison has its specific target period. Thus, there are several predictions corresponding to various initial values $N(t_0)$.

Figure 11 shows results of the prediction for the 1980 population estimate. The number of 9-year-olds is obtained by the back projection procedure as described above. The target period for the prediction is between 1970 and 1990 with a center at 1980. One can observe an almost precise prediction near 1980. Before 1975 and after 1985, the curves diverge. For the period before 1975 this behaviour can be explained by a growing with time number of people for the same birth year cohort. Obviously, due to a positive international migration into the USA, the number of 9-year-olds in 1960 was lower than the number of 29-year-olds in 1980. Following the same logic, one can expect a larger predicted number of 9-year-olds than the projected one for younger population.

Figure 12 compares 1990 the population estimates. The years around 1990 are described not so successfully as those around 1980 and 2000. There is a notable oscillation in the estimated number, however, that mimics to some extent the predicted behaviour. The worst period for the prediction is between 1984 and 1988, where the discrepancy is the largest and the predicted curve even goes out of the 5% uncertainty bounds. The 1990 population estimate is very well matched by the predicted curve for the period before 1980. Despite an obvious underestimate in the population level the predicted curve gives an excellent description for the shape of the estimated curve. In addition, the predicted curve shows an excellent agreement with the observed one towards 2000.

Figure 13 and 14 present similar comparison of the counted and predicted from the observed GDP number of 9-year-olds. The counted number is obtained as a projection of the 2000 population estimates – the post- and intercensal one, respectively. The predicted curve almost coincides with the postcensal population estimate during the period between 1996 and 2001. The intercensal estimate corrected for the 2000 census enumeration also provides a good approximation for this period, but does not contain a small peak near 2002 and is characterized by a steeper drop of the 9-year-olds number after 2001.

Figure 15 illustrates a comparison of the predicted curve with the reported birth rate in the USA shifted by 9 years ahead. The birth rate curve, however, is also prone to modification by migration processes during the 9 years. General features of the birth rate curve are well predicted and only the observed population level is higher than predicted.

Figure 16 shows the predicted and observed 9-year old population. The population estimates (post- and intercensal) resulted from numerous modifications and updates related to the decennial censuses and intercensal population estimates. These estimates also include a number of corrections related to some independent methods of population estimates. It is the result of the best efforts of the Census Bureau to present a consistent population universe which is aimed to balance all available information and theoretical knowledge. Hence, one can assume that the requirement of a balanced population universe contradicts accuracy of the single year population estimates. Nevertheless, the 9-year old population predicted from the observed real GDP describes well the enumerated population of this age.

Concluding this section, we briefly summarize our findings. The best agreement between the estimated and predicted number of 9-year old population is observed for the projected decennial census population from 3 to 5 year around the year of enumeration. Migration processes do not disturb the population age distribution much. In the years between the censuses, the curves diverge the most. The population estimates for these years are obtained by the component method relied on independent measurements of the birth and mortality rates and migration dynamics. All these components might be inaccurately measured and/or artificially biased. The methods for correction of the single year of age population distribution are very likely to be inaccurate as follows from the comparison of the postcensal and intercensal values. The methods do not consider any possible fluctuation of a single year population between the census years and just distribute proportionally "the closure error" over the years. One can assume that any curve with the same values at the census years can fit the observations, including the curve which is obtained from the real GDP observations according to relationship (3). This is the principal conclusion an the answer to the second question of section 3 – the GDP predicted curve does not contradict the observed population estimates considering their accuracy and methods of correction and modification.

The predicted curve contains two principal components – a high frequency one related to the 9-year old population change and a trend component. The latter is consistently subtracted from the observed GDP growth rate. The component depends on the evolution of the value measured by independent methods, the $T_{cr}$.

Thus, the agreement between the predicted and observed population supports the hypothesis of the dependence of the real GDP evolution on the two measurable parameters. The better one knows the 9-year-old population distribution, the more accurate is the prediction of the real GDP evolution. Theoretically, it is possible to measure the population without any uncertainty as the number of people is an integer value. In turn, the GDP growth rate prediction will be as precise as the population measurement.

## 5. The United Kingdom and France

For any economic theory it is important to be a universal one and applicable to similar objects. There are several large developed countries providing information on population age distribution and real GDP growth rate. The GDP prediction from population and the reversed procedure is accomplished for developed countries as well. Here we present result for the UK and France. Figures 17 through 25 show the GDP and population predictions in the same order as those for the USA. The general conclusion is that the observations do not contradict the theoretical approach developed for the USA. One can observe very good correspondence between the deep recessions in the beginning of 1980s and 1990s and the lack of 9-year-olds in the UK. The same is valid for France near 1993. The only principal difference between the UK and France is that the defining age for France is eighteen years. This age occurred to be the defining one also for Japan, Germany, Austria, and Russia – the countries studied by the author who has no explanation of the age difference. Accuracy of the population estimates for the UK and France is hard to estimate, but some features unveil some artificial character of the population age pyramid construction.

## 6. Per capita GDP

In previous sections we have developed a model explaining the observed real GDP growth rate variations in past. We have linked the GDP observations with two principal parameters – duration of the mean income growth period and the number of 9-year-olds. There is no explanation of these relationships. The model is an empirical one and do not pretend to explain individual and/or social behavior resulting in increase of the period of the mean income growth with increasing per capita GDP. Economic life of a country is a complicated system of interactions between people which lead to a hierarchical and rigid structure of personal income distribution as found by Kitov (2005a). It was shown that any change in population age structure and real economic growth both induce predefined changes in the corresponding PID, i.e. there is a functional dependence between these parameters. They create a dynamic economic system similar to those observed in natural sciences.

There is another parameter characterized by a potential predictability for an evolving economy. This parameter is the absolute value or annual increment of per capita GDP growth. Similarly to the previous sections, one can distinguish two principal sources of per capita GDP growth: the change of 9-year old population and the economic trend related to per capita values. The trend has the simplest form – no change of the mean annual increment. This is expressed by the following relationship:

$$dG/dt=A \quad (4)$$

where $G$ is the absolute level of GDP per capita, $A$ is a constant depending on country. The solution of this equation is as follows:

$$G(t)=At+B \quad (5)$$

where $B=G(t_0)$, $t_0$ is the starting time of the studied period. Relative growth rate is then expressed by the following simple relationship:

$$dG/Gdt=A/G \quad (6)$$

which indicates that the relative growth rate of per capita GDP is inversely proportional to the GDP per capita and the observed growth rate should decay in time with asymptotic value equal to 0.

Figure 26 represents approximations of the observed GDP per capita annual increments by a constant for the USA ($485 – 2002 US dollars), France ($405), Italy ($405), Japan ($480), and the UK ($378) (CB and GGDC, 2006). The constants are obtained as the mean value of the absolute growth between 1950 and 2004, i.e. for the last 55 years. For the USA, a longer period is also presented with its own estimate of the growth constant. One can see that the values of the mean growth are between $400 and $500. Only the USA and Japan have a larger value among the largest economies. Germany is not present due to the change the country underwent in 1990s. Smaller economies also are not analyzed in this paper because their economic behavior might be biased by some factor of external origin. For example, presence of some large neighboring

economies making the production of the small country oriented mostly to the demand of the large-size neighbor.

Another possibility is a development of a monopoly service or production driving such a small economy for a short or longer time period. In all these cases, the small economy can be considered as an industrial or service part of a bigger economy and can not be sustainable by itself. An analog of a small economy can be an industrial branch in a large economy that can potentially develop at a rate different from all other industry branches, i.e. increasing or decreasing its part in the total production, but can not be independent. Thus, economic growth of a small economy might not be described by the same relationship as that for a large economy, where effects of fast increase in one sector of economy are usually compensated by decrease in another sectors.

When applied, a linear regression (corresponding equations for the linear regression are given for the countries in Fig. 26) gives practically the same constant line for France, Italy, and Japan, i.e. the best approximation of a linear growth is just a constant for these countries. For the USA and UK, the linear regression indicates some positive trend in per capita GDP increment with increasing GDP per capita. As shown later, this increase might be related to the effect of the population component change. For the USA, one can presume a downward trend in future, when and if the effect of the population component of growth will disappear. Such a behavior was clearly observed in Japan in the 1960s and 1970s, when the population effect was very strong and the per capita GDP absolute growth increased. The effect has being compensated during the last 20 years due to a severe drop in birth rate. One can expect the trend to continue in Japan because of the negative influence of the population change. If the constant of the absolute growth in Japan is the same as in other four countries, i.e. ~$400 as shown below, then the accumulated population effect is of $80.

Figure 27 summarizes the mean GDP increment values for some developed countries. One should bear in mind, however, that the mean values include some input from the both sources of growth. The population change factor (relationship (1)) should be subtracted from the total absolute growth value. For example, the 9-year old population in the USA changed from 24022326 in 1950 to 4173171 in 2002. The total input of the population change in the observed economic growth is 0.5*(4173171-2402326)/ 2402326=0.37. Per capita GDP grew during this period from $12123 to $38345 or by a factor of 2.16. The total growth can be split into the population related component of 0.37 times and the economic trend component of 2.16-0.37=1.79 times. The mean GDP per capita growth was \$485 and includes \$401=\$485*(1.79/2.16) of the constant absolute growth (economic trend) and \$84=\$485*(0.37/2.16) of the population related growth.  For France this effect is much weaker than for the USA – the number of 18-year-olds changed from 649011 in 1950 to 801095 in 2001, i.e. the population effect for the observed growth is only 0.12 or 3 times lower than in the USA. The total growth in France was from \$7009 to \$28956 or 4.13 times (\$406 per year in average) from which only 100*(0.12/4.13)=2.9% were related to the change in the specific age population. When corrected for the population change, the observed mean per capita GDP growth in France is almost the same as in the USA: \$394=0.97*\$406. The same procedure can by applied to other countries when corresponding data is available.

Another principal correction has also to be applied to the per capita GDP values. This is the correction for the difference between the total population and population of 15 years of age and above as discussed by Kitov (2005a). The concept requires that only this economically active population should be considered when per capita values are calculated. Relevant data is not available for countries except the USA and France. We use the published (uncorrected) per capita GDP values bearing in mind potential bias induced by the discussed population effects.

Figure 28 illustrates the per capita GDP relative growth rate as defined by relationship (6). The observed values are shown together with some economic trend curves obtained by a regression with a function *A/G* (where *A* is the absolute value of constant growth; *G* is the GDP per capita). Black and red curves correspond to the mean increment between 1930 and 2002 (A=\$399) and between 1950 and 2002 (A=\$535), respectively. The regression is conducted not by the least squares because this standard method does not preserve the total observed growth during the period, i.e. the sum of all growth rate values for the observed GDP values. We used a regression which minimizes the overall deviation of the observed relative growth values from the best-fit function *A/G*. Such a regression preserves the total observed growth for a given period with the observed values of per capita GDP.  The relative growth rate is a poor parameter to describe economic growth for a relatively long time interval. The relative growth rate is calculated with a changing baseline value. One can not even sum (or multiply) the relative growth rates in order to obtain some reasonable value of the total growth. For example, an averaged over 50 year long interval relative growth rate of 2% (i.e. mean value of 50 independent measurements) gives a lower absolute growth than an averaged  rate of 1.9%  if the amplitude of oscillations around the averaged value in the first case is 5 time larger than in the second. Figure 29 illustrates this fact - two curves correspond to oscillation amplitude of 0.05 around the mean growth rate of 1.02 and of 0.01 around the mean growth rate of 1.019 during 50 years. Formally, the average growth rate of 1.02 results in lower total growth of 2.53 than that for the average growth rate of 1.019 reaching 2.56 at the end of the period. Thus averaging of relative growth rates over long intervals is not a correct procedure because it depends on the growth rate oscillation amplitude. Absolute growth characterizes evolution of per capita GDP in an appropriate way because the averaging operation is correct.

Magenta curve in Figure 28 illustrates the relative growth related to the economic trend only, i.e. to the total economic growth less the population component. As found above, this value is approximately 1.79/2.16=0.83 of the observed mean absolute growth between 1950 and 2002. The fluctuations in the growth rate values around this curve has to be explained by the population factor. Here and below all the per capita GDP values are corrected for the difference between the total population and the population of 15 years old and above. The ratio of the populations in the USA changed from of 1.37 in 1950 to 1.27 in 2002. In 1930, the ratio was 1.41. Thus the per capita GDP values are larger by 1.37 to 1.27 times than in the original table.

One can rewrite equations (1) and (3) in order to replace real GDP with per capita GDP and also incorporate economic trend as defined by relationship (6):

$$g_{pc}(t)=0.5*dN(t)/N(t)dt+A/G(t) \quad \textbf{(7)}$$

where $g_{pc}(t)$ is the per capita GDP growth rate, $G(t)$ is the real per capita GDP as a function of time. One can obtain an inverse relationship defining the evolution of the 9-year-old population as a function of economic variables:

$$d(lnN(t))=2(g_{pc}-A/G(t))dt \quad \textbf{(8)}$$

Usage of relationships (7) and (8) is similar to that of relationships (1) and (3). Figure 30 presents results for the per capita GDP growth rate prediction by using the 2000 postcensal population estimate according to relationship (7). As previously, the number of 9-year-olds was derived by the projection of the age structure back and forth in time. As in the case of the total GDP prediction, the principal features of the observed curve are in a good agreement with the predicted ones, including severe drops in per capita GDP during the years of recession. There is no possibility to reach a better agreement because of the limited accuracy of the population estimates compared to that for the per capita GDP.

Figures 31 through 34 display results of the prediction of the number of 9-year-olds from the observed per capita GDP according to equation (8). The figures are similar to those for the total real GDP. The best fit is obtained for the number of 9-year-olds as estimated by the Census Bureau in the postcensal estimation procedure. The curves almost coincide between 1960 and 2002. As discussed above, the estimated curve is slightly over smoothed compared to the census population estimates.

Hence, the observed behavior of per capita GDP growth rate can be successfully described by the model of two sources of growth. If to replace the estimated population (red curve) by the predicted population (blue curve) one can reach the absolute prediction accuracy. The specific age population change has a wavelength of 30 years. Accordingly, the GDP growth rate reveals the same feature merged with the decreasing economic trend. Some oscillations with shorter periods are also clear. All in all, the presented model for the economic growth does not contradict the observations and has a predictive power that only depends on the capability to accurately count people of some known age.

Results obtained by the same procedure applied to France are presented in Figure 35. This comparison is similar to that presented in Figure 25 because the influence of the population component is just minor according to our estimates of the total change of the number of 18-year olds between 1950 and 2002. In the long run, the per capita GDP and total GDP depend only on inherent growth of the French economy.

## 7. Discussion and conclusions

We have presented an empirical model predicting real GDP behavior by using only two parameters – per capita GDP and the number of people of some predefined age. There is an important difference, however, in economic trend for the total real GDP and per capita GDP. The latter is related to the (observed) constant absolute growth and the corresponding relative per capita GDP growth rate inversely proportional to the per capita GDP itself. The former is related to the (observed) evolution of some critical work experience, $T_{cr}$, inversely proportional to the square root of the per capita GDP. In the frame of this study, one can not decide which of the two parameters is defining – the value of the absolute growth or the evolution of the $T_{cr}$. To distinguish between the two effects one need more data on the $T_{cr}$ evolution in different countries. Formally, it is possible to obtain the observed total GDP economic trend to be inversely proportional to the square root of the per capita GDP, if to add the observed change of the total working age population to the predicted change of the per capita GDP, which is inversely proportional to per capita GDP.

**Figures**

**a)**

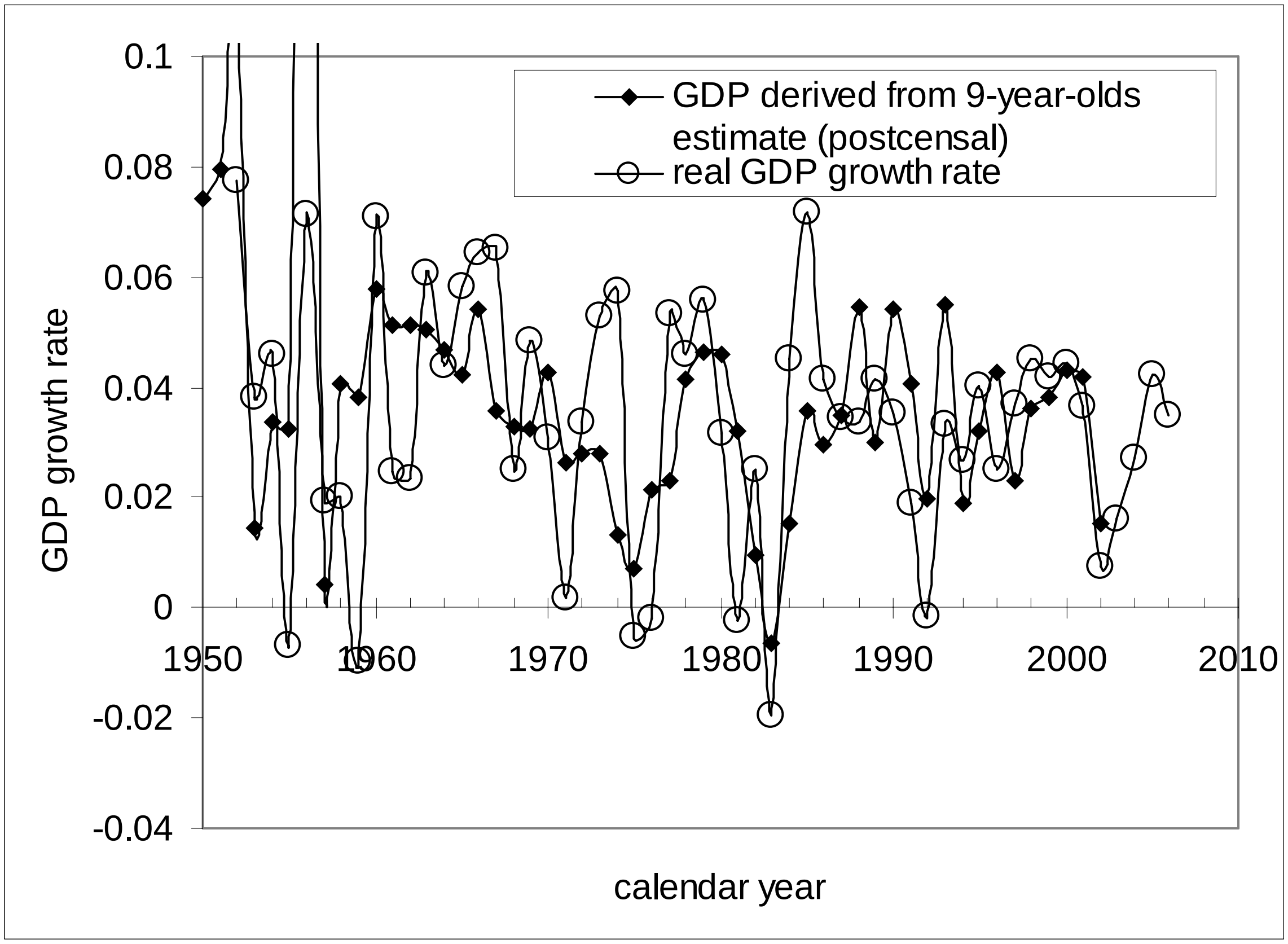

**b)**

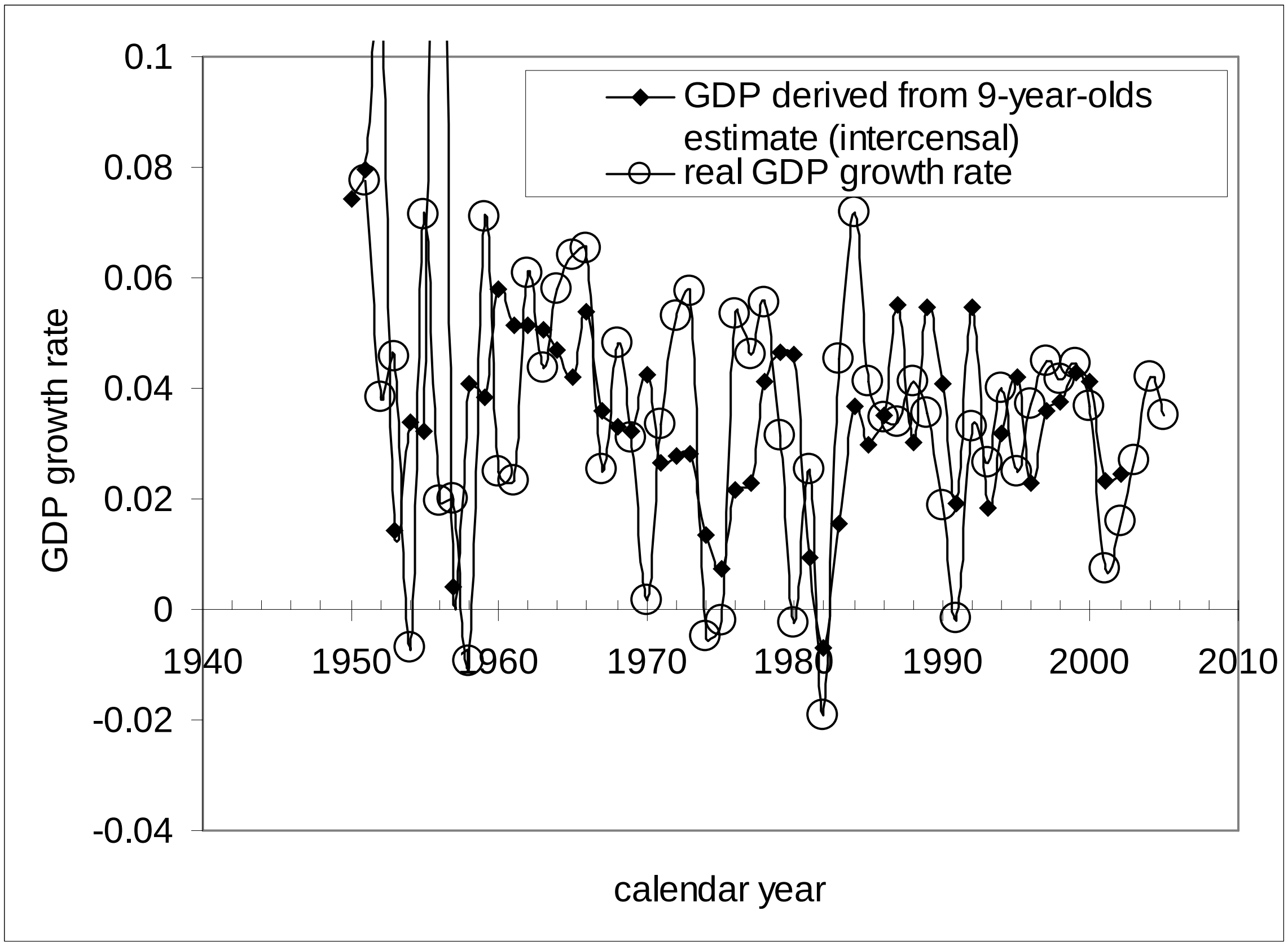

Fig. 1. Real GDP growth rate in the USA for the period between 1950 and 2004. Comparison of the measured values and those predicted from the single year of age population estimates (9-year-olds): a) 2000 postcensal estimates; b) 2000 intercensal estimates (US CB, 2005a-d ).

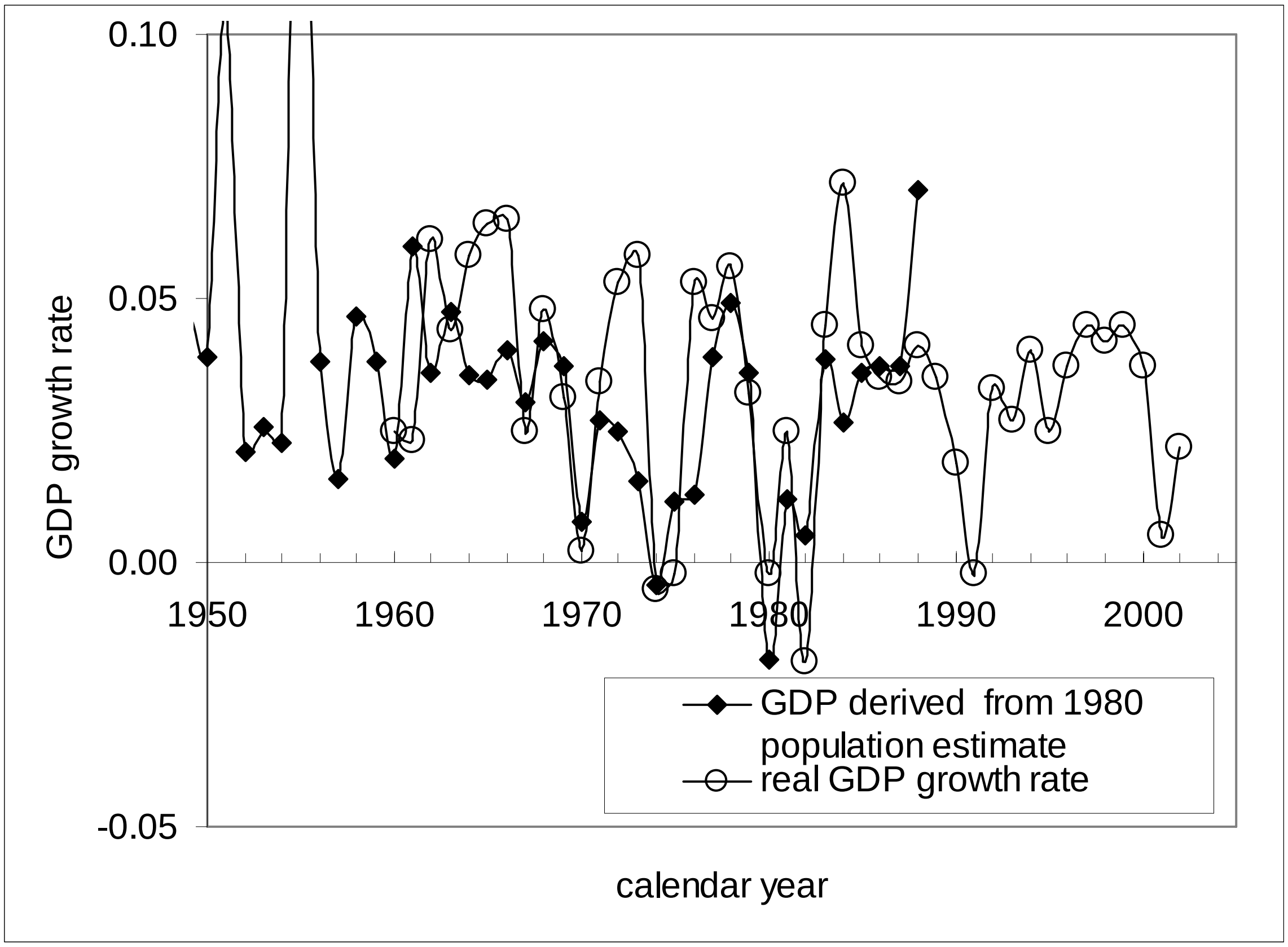

Fig. 2. Real GDP growth rate in the USA for the period between 1960 and 2002.Comparison of the measured values and those predicted by the population estimates of 9-year-olds made as a projection of the 1980 single year of age population estimate into the 9-year-olds time series. Notice the predicted and observed curve behaviour around 1980 - oscillations are synchronized. The pattern is smoothed in the subsequent vintages and is not observed in the distribution of the 9-year-olds.

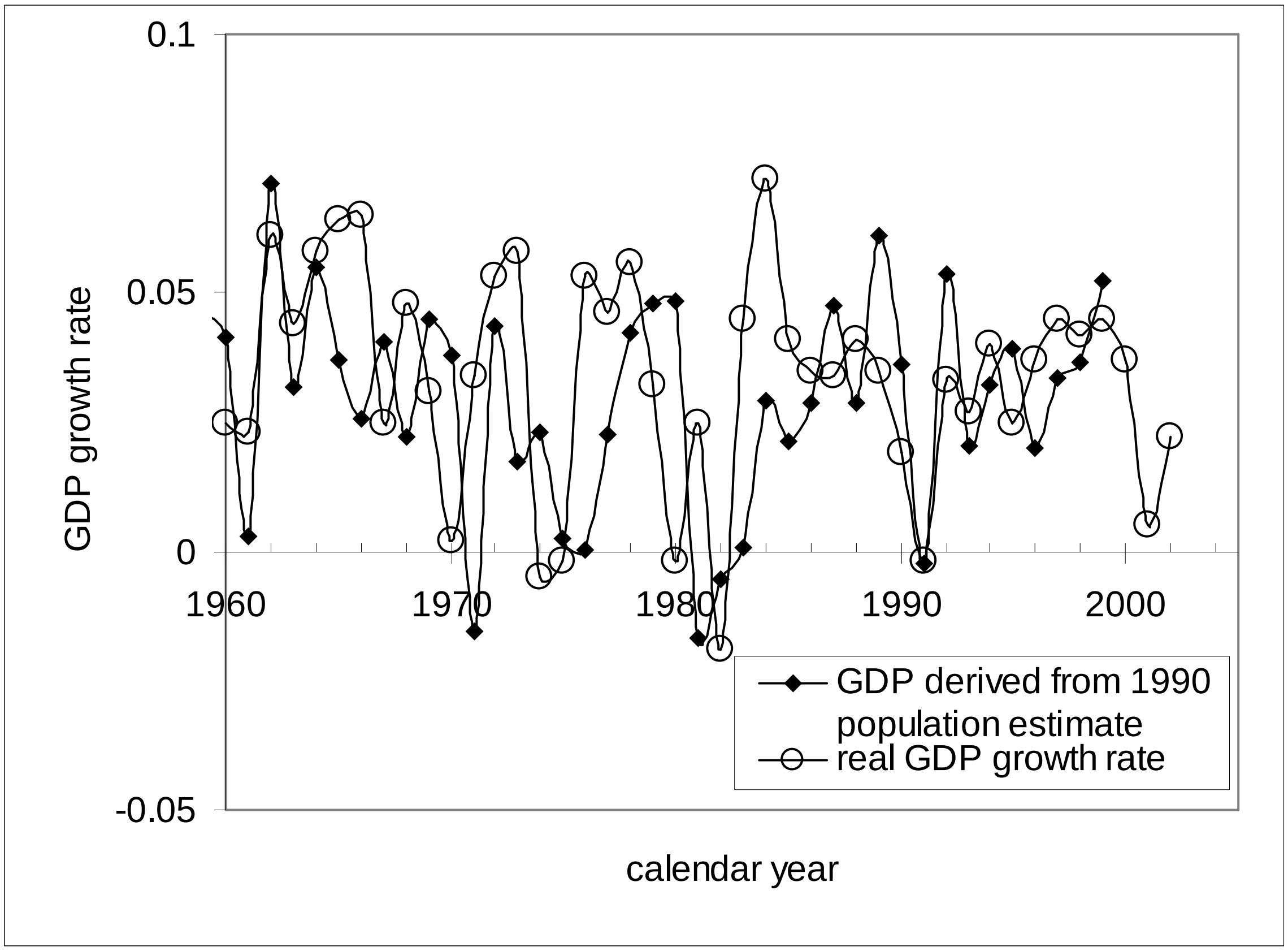

Fig. 3. Real GDP growth rate in the USA for the period between 1960 and 2002.Comparison of the measured values and those predicted by the population estimates of 9-year-olds made as a back projection of the 1990 single year of age population estimate. Notice the predicted and observed curve behavior around 1990. The fluctuations observed near 1980 in Fig. 2 are not observed any more. The oscillations correspond to the ages between 17 and 22 in the 1990 estimate.

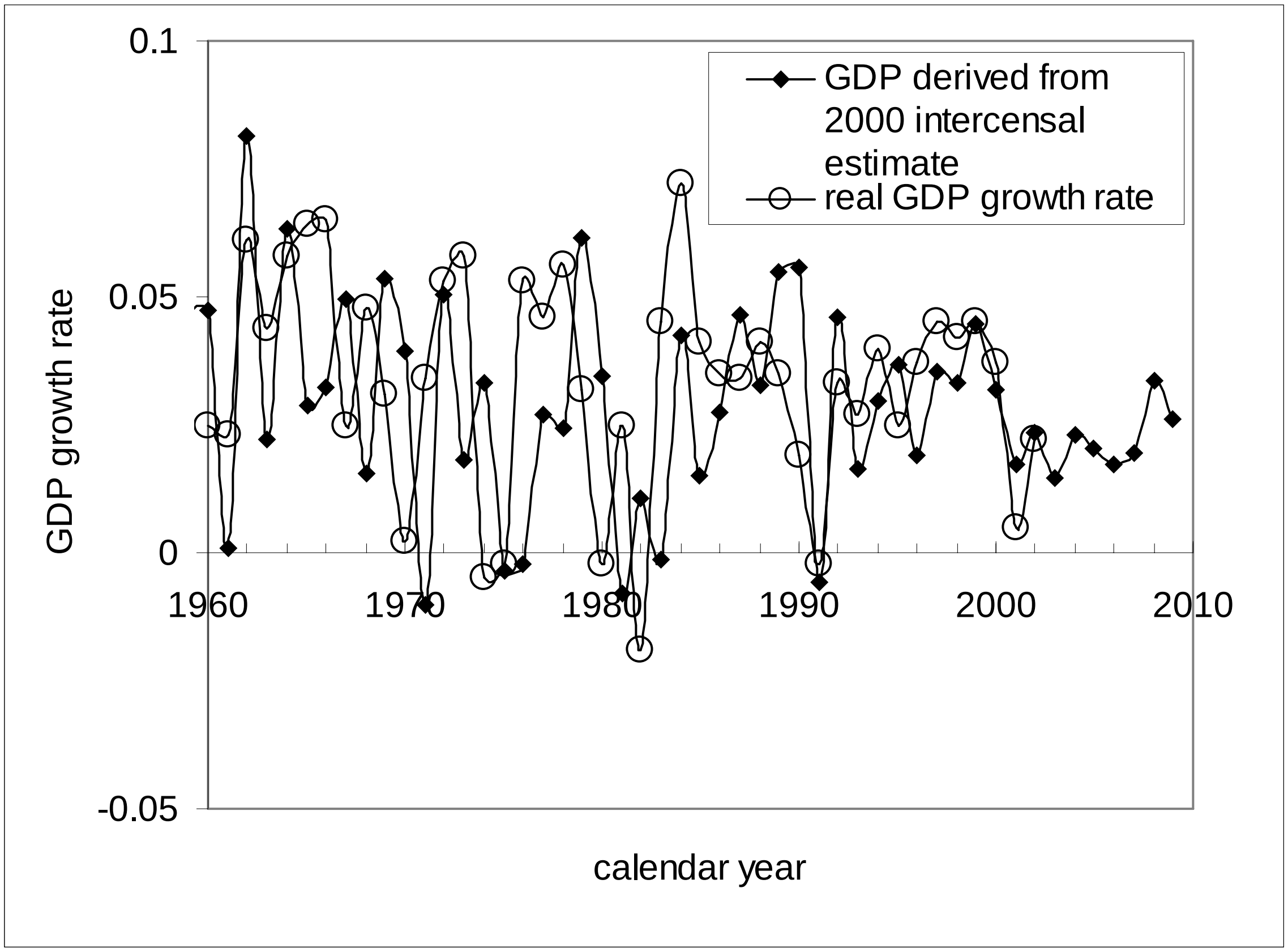

Fig. 4. Real GDP growth rate in the USA for the period between 1960 to 2001.Comparison of the measured values and those predicted by the population estimates of 9-year-olds made as a projection of the 2000 single year of age population ***intercensal*** estimate. Notice the predicted and observed curve behavior around 1980, 1990, and 2000.

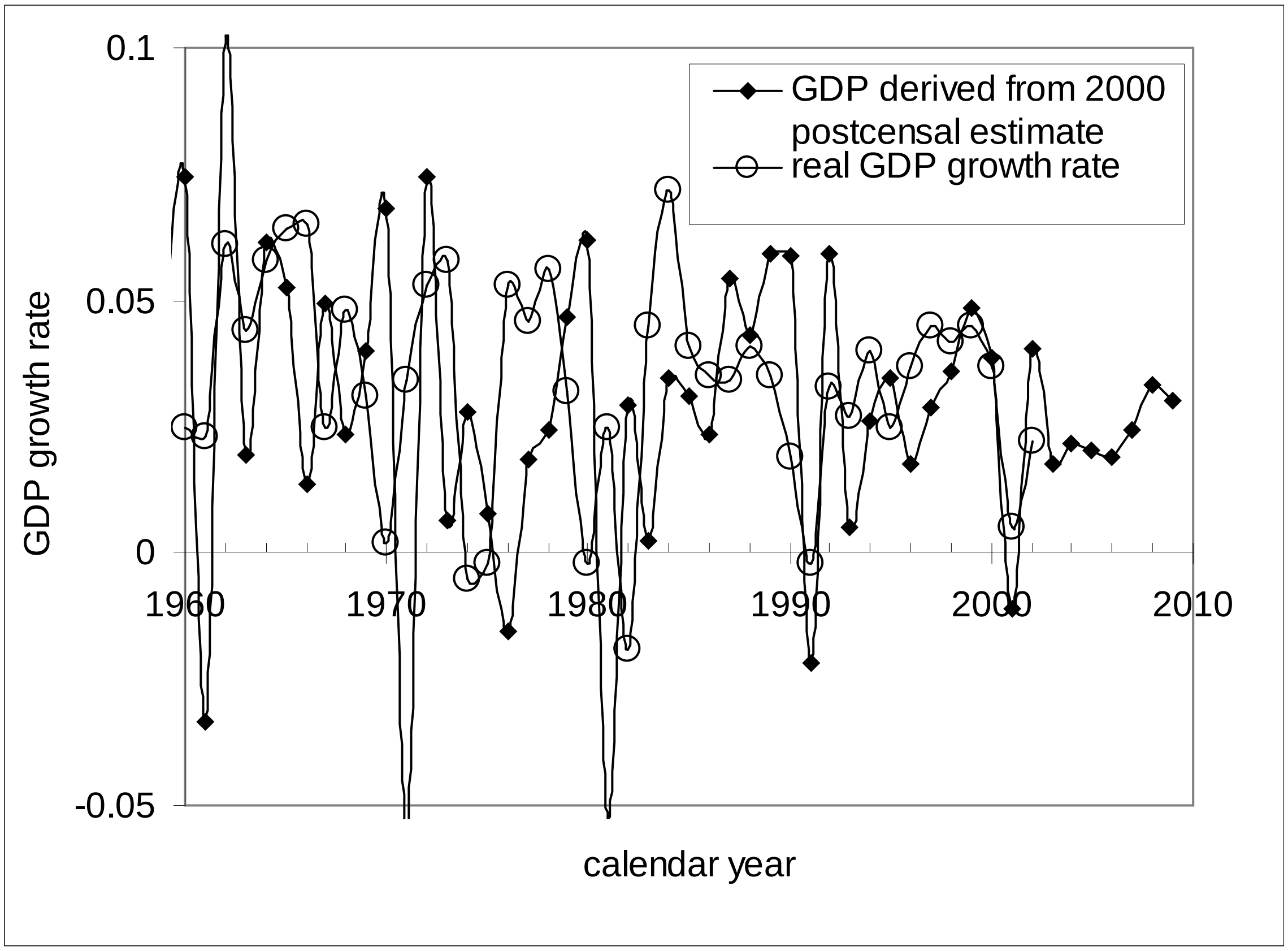

Fig. 5. Real GDP growth rate in the USA for the period between 1960 and 2001.Comparison of the measured values and those predicted by the population estimates of 9-year-olds made as a back projection of the 2000 single year of age population ***postcensal*** estimate. Notice the difference with the intercensal estimates shown in Fig. 4.

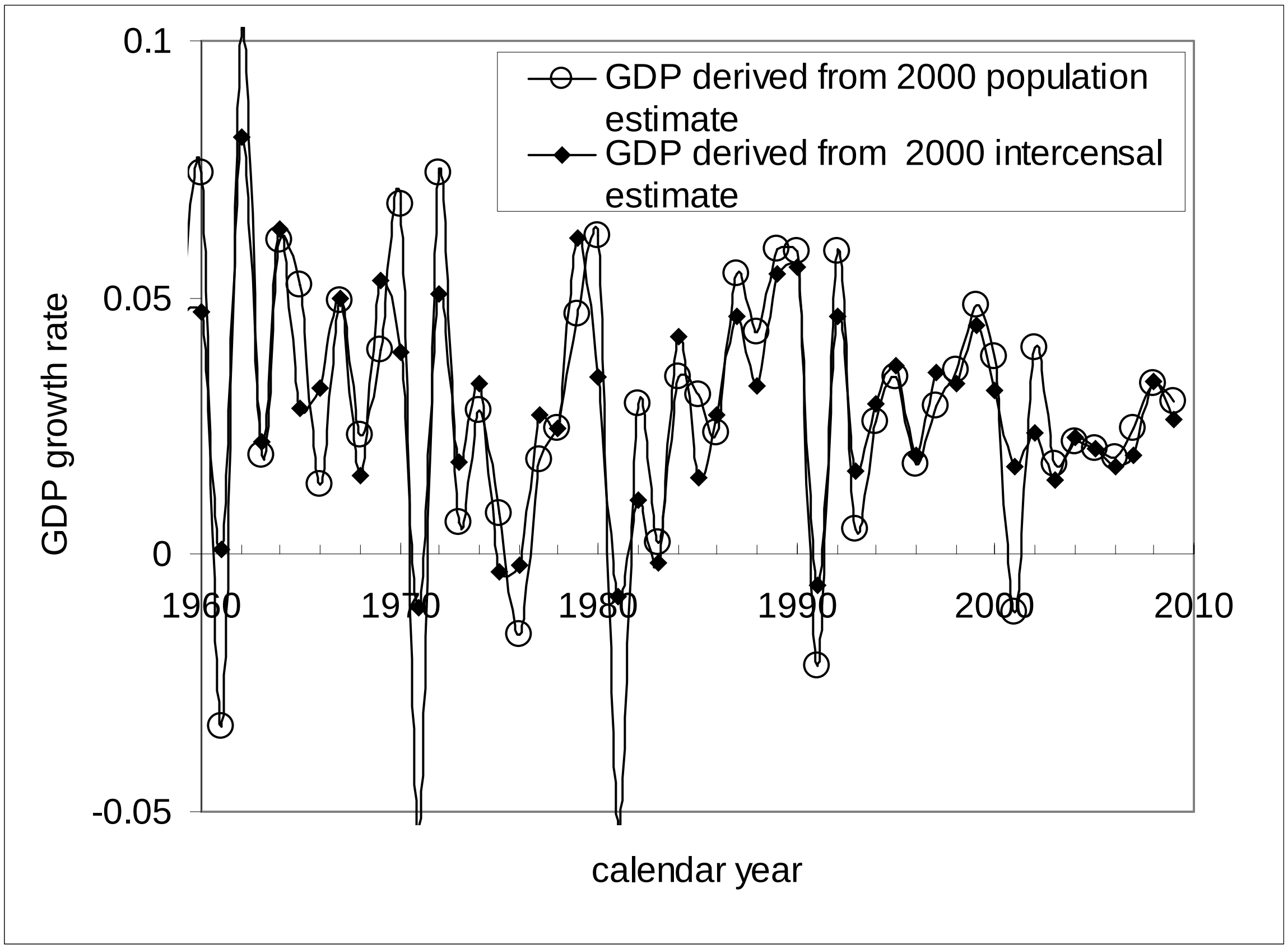

Fig. 6. Predicted GDP growth rate in the USA for the period from 1960 to 2001.Comparison of the values predicted by the population estimates of 9-year-olds made as a back projection of the 2000 single year of age population ***postcensal*** and ***intercensal*** estimates. The latter is a severely smoothed version of the former.

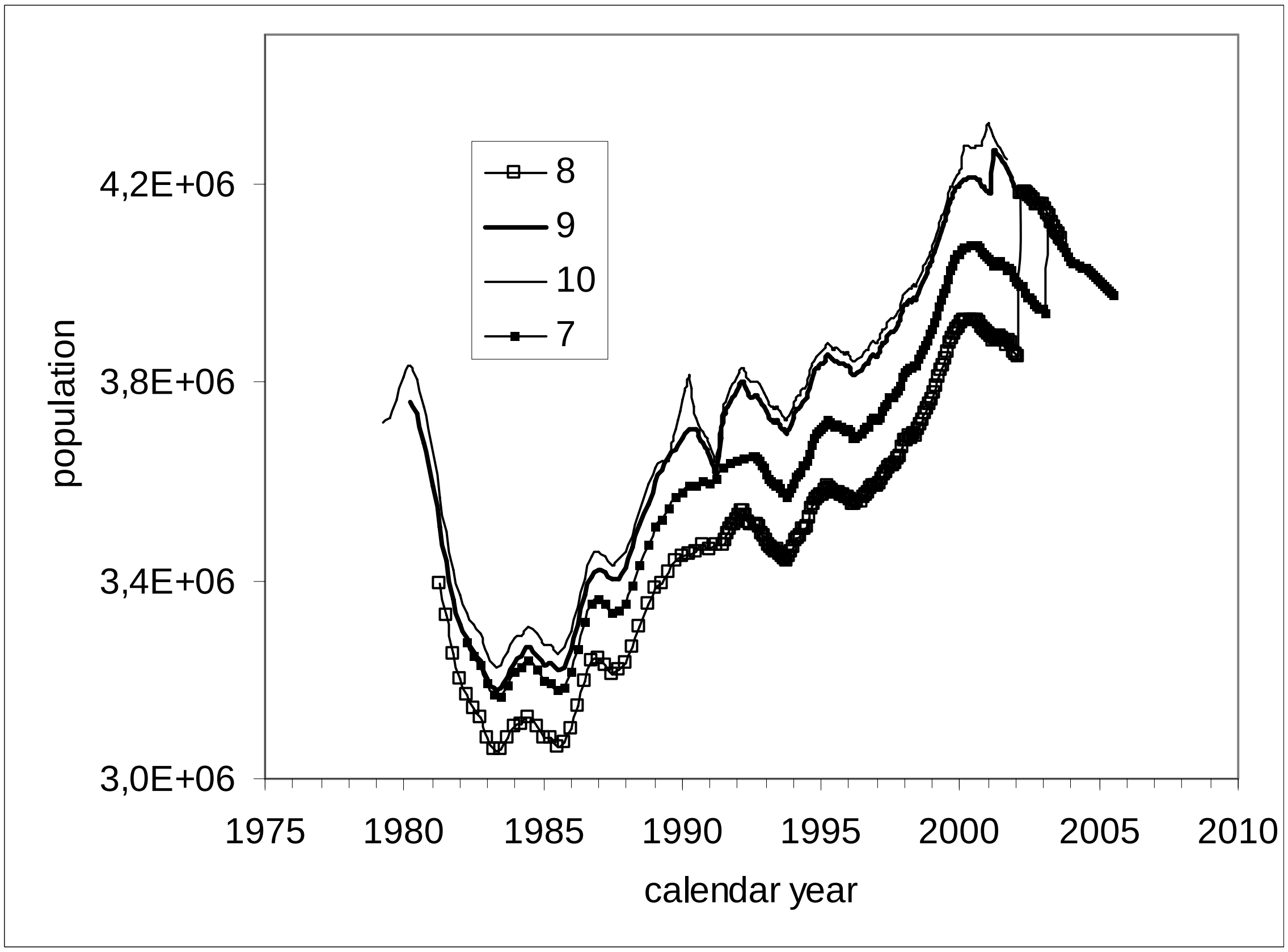

Fig. 7. Comparison of the quarterly (before 1990) and monthly (after 1990) ***postcensal*** estimates for several single year of age populations: 7, 8, 9, and 10 years. The distributions are shifted in time relative to the 9-year-olds curve in order to trace the same cohort. Notice the difference between the 8-year-olds and 9-year-olds reaching 300,000. The number of 7-year-olds is consistently above the number of 8-year-olds violating the general rule of increasing cohort population.

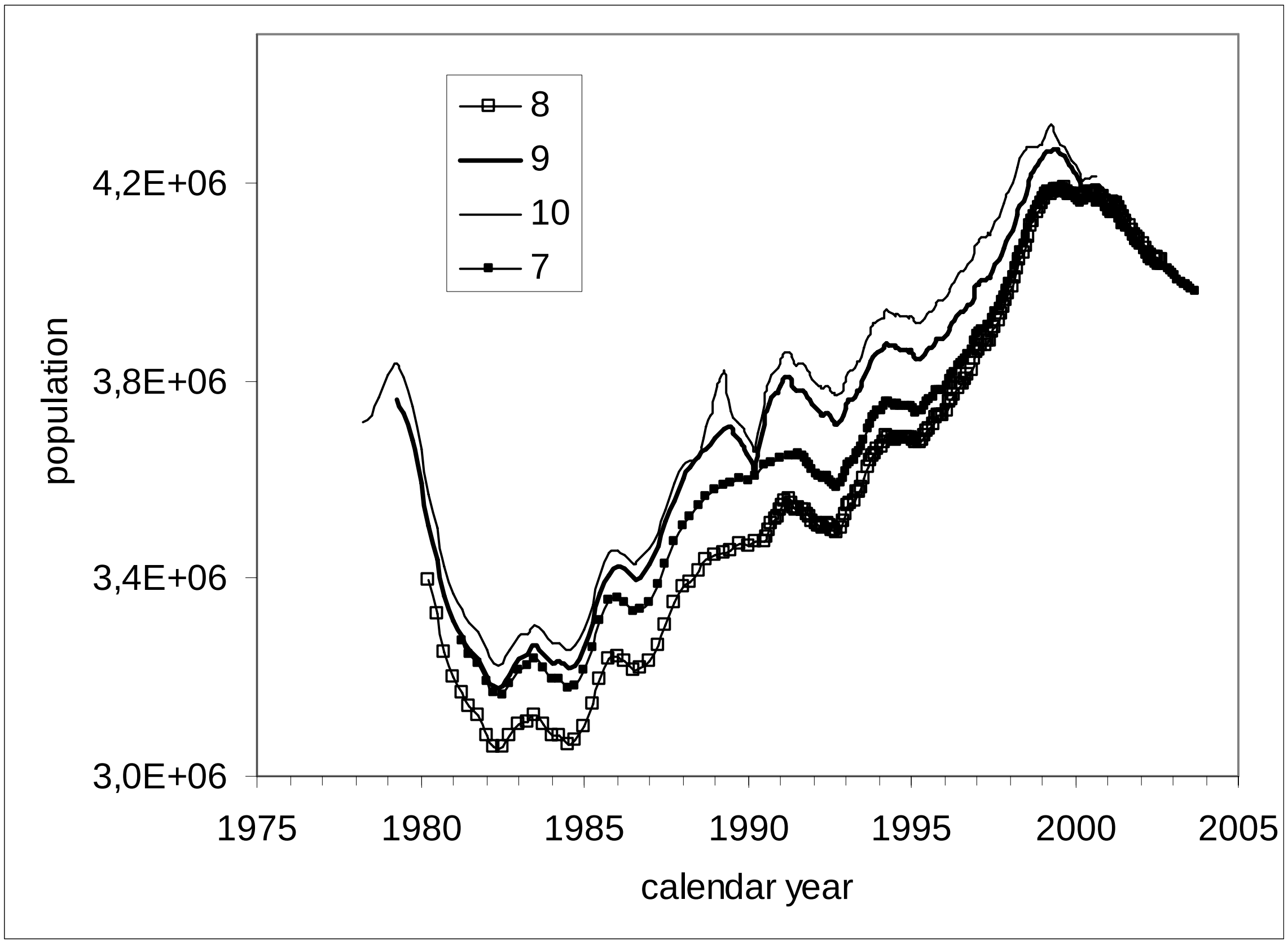

Fig. 8. Same as in Fig. 7 for the ***intercensal*** (1980-2000) estimates. Notice the convergence of the curves for the 7-year-olds and 8-year-olds as compared to with the postcensal estimates. This correction gives an approximate estimate of the uncertainty of the 9-year-olds number.

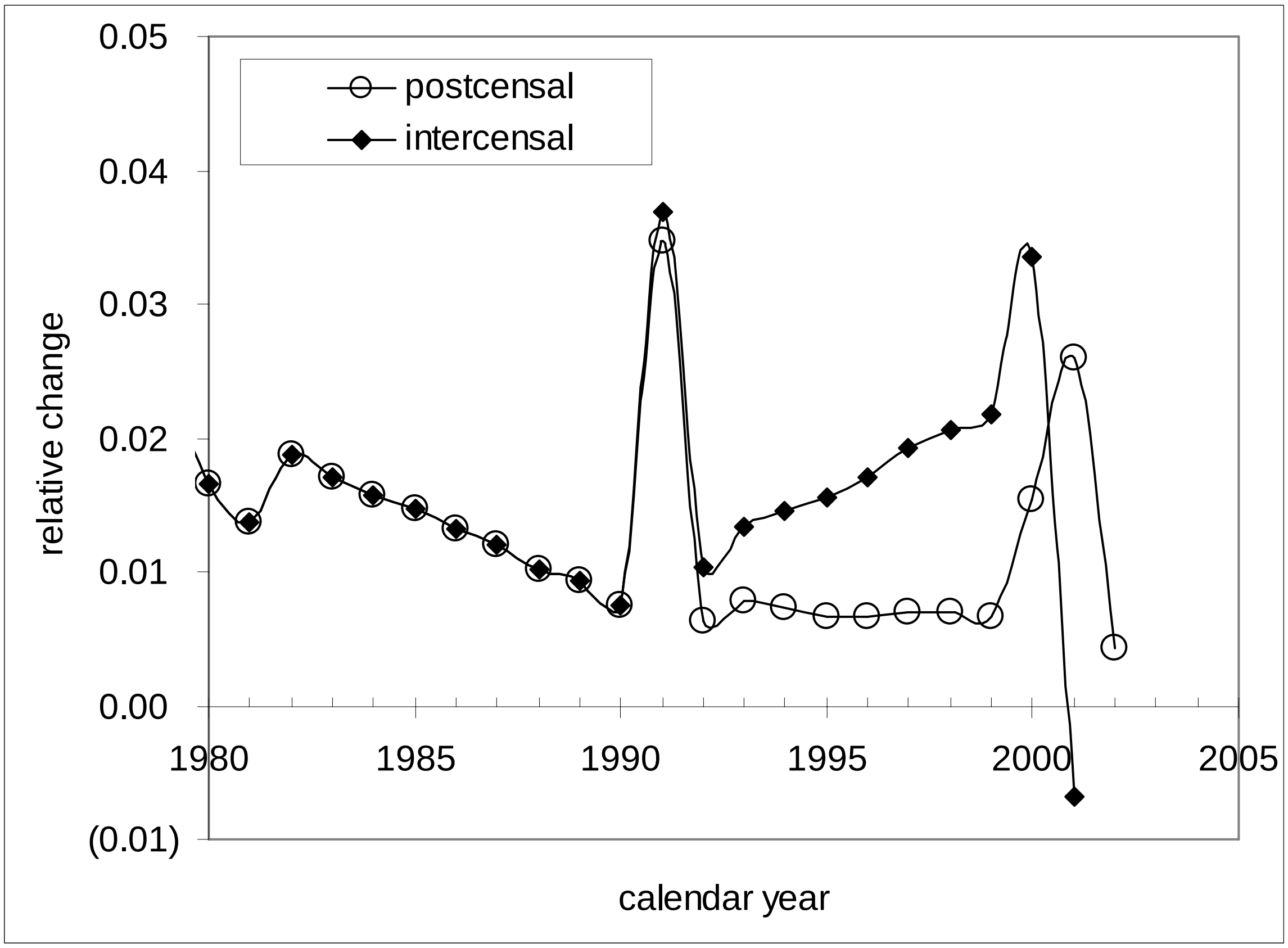

Fig. 9. The change of the 9-year-old population during one year, i.e. the difference between the 9-year-olds and 10-year-olds in the next year. Notice severe population changes near the census years and linear growth of the difference between the censuses.

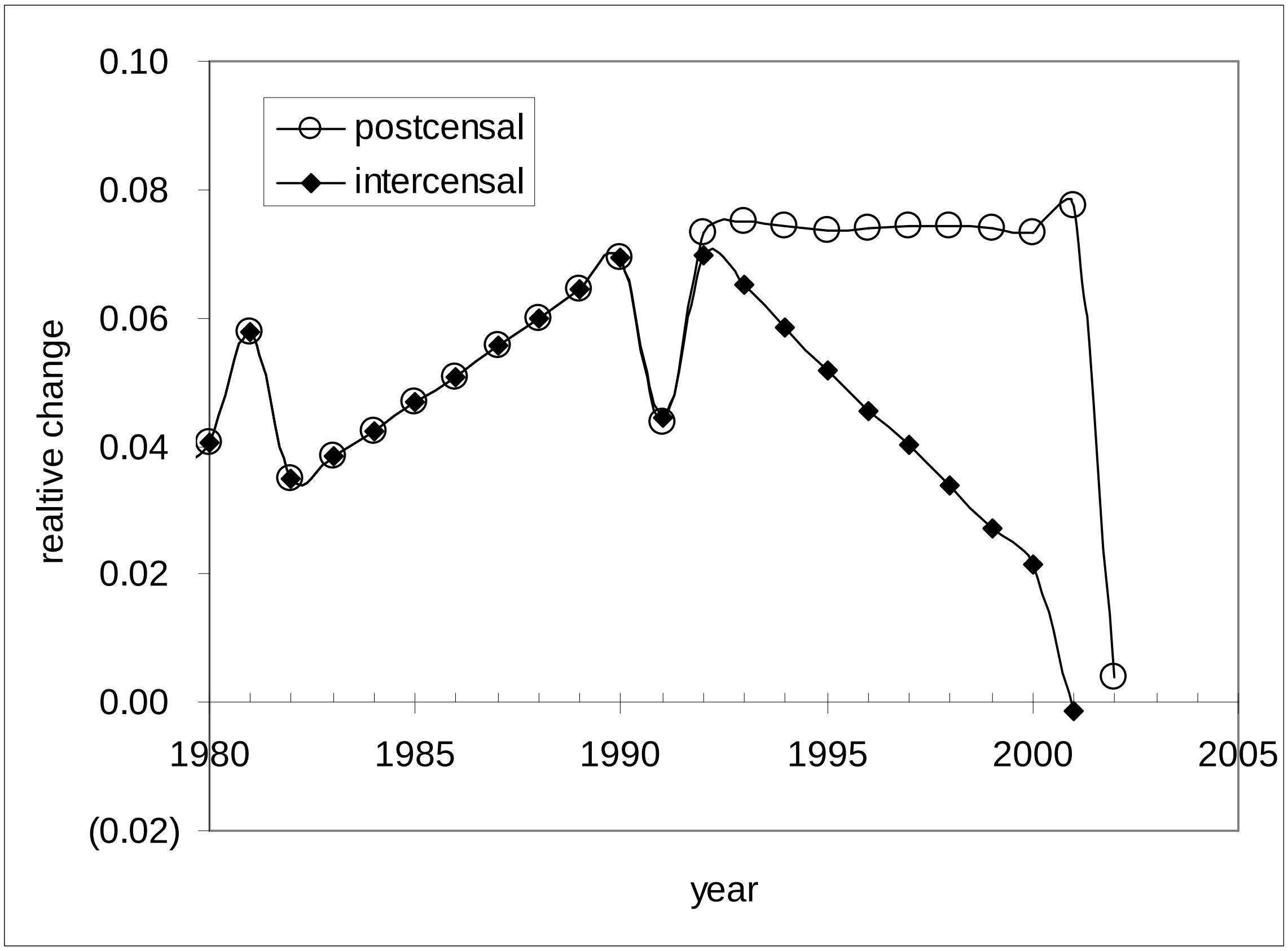

Fig. 10. Same as in Fig. 9 for the 8-year-olds. Notice the two corrections in 1981 and 1991. One can suppose the uncertainty of the 9-year-olds number of several percent.

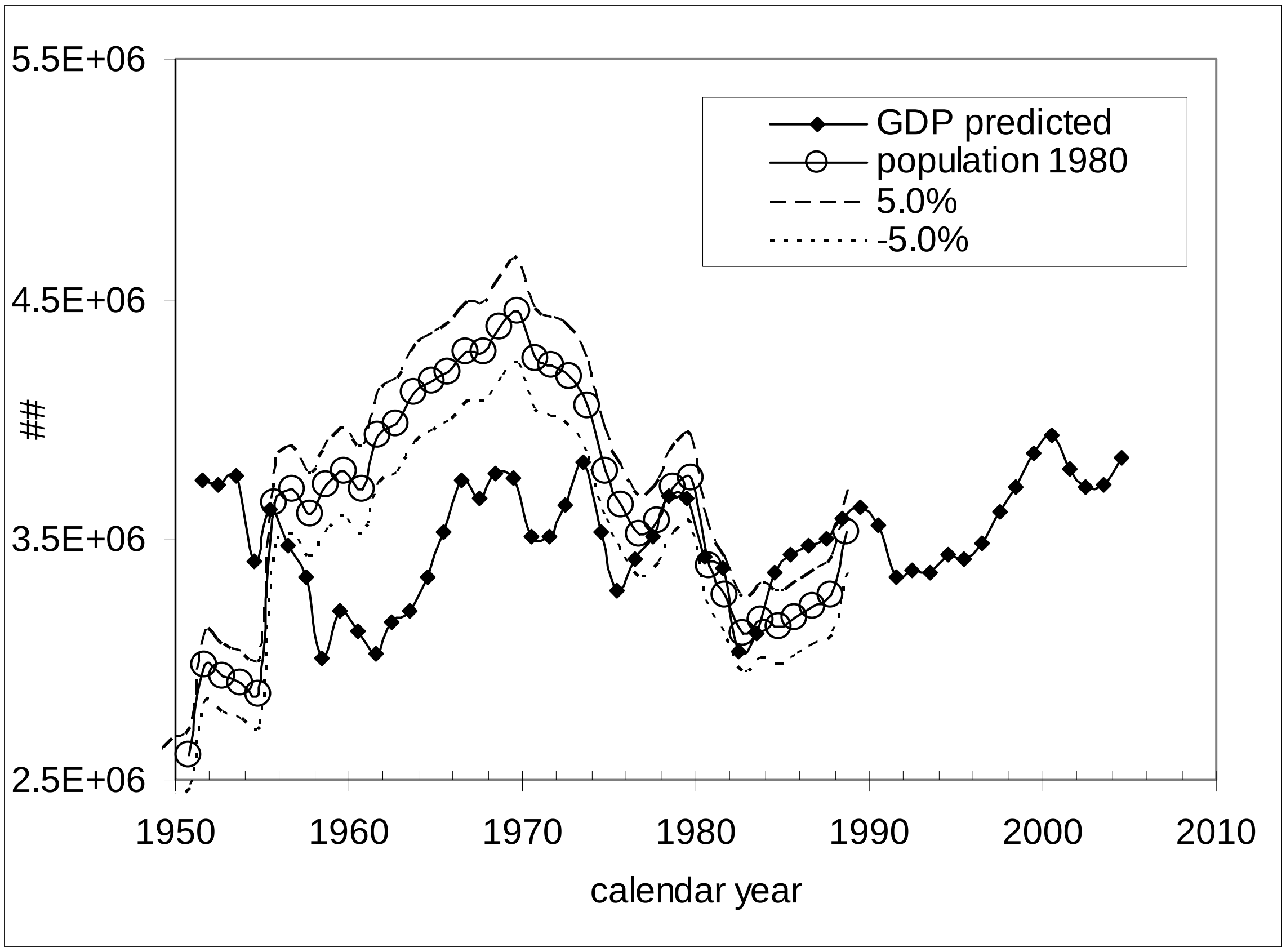

Fig. 11. Comparison of the 9-year-old population as obtained by a back projection from the 1980 population estimate and the number of 9-year-olds obtained from the real GDP observations according to equation (3). Notice the coincidence of the curves near 1980. The initial value of the 9-year-olds in 1951 is 3750000. Actual number of the 9-year-olds is obviously overestimated by the projection for the period before 1975 due to a positive demographic dynamics in the USA.

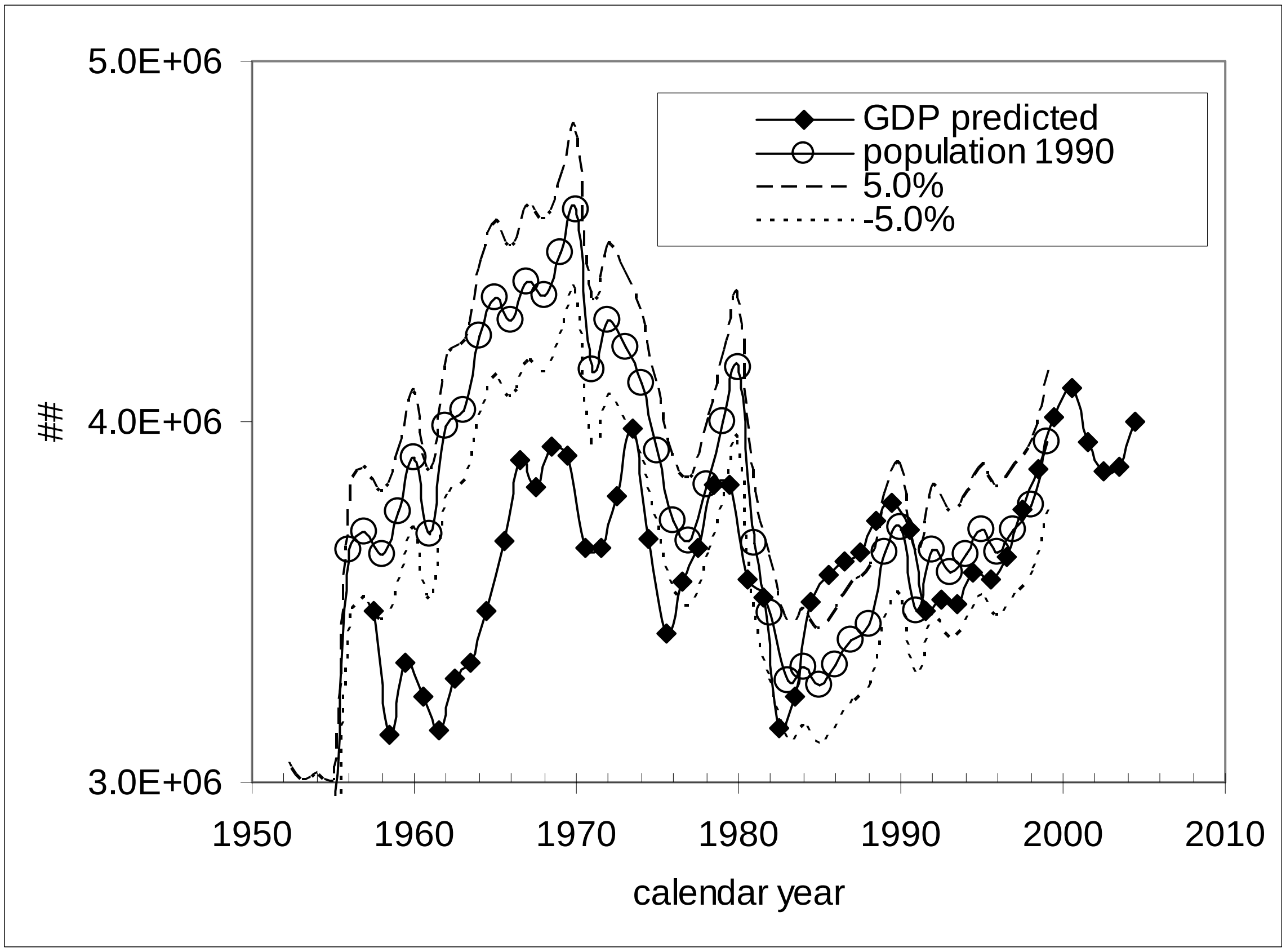

Fig. 12. Comparison of the 9-year-old population as obtained by a back projection from the 1990 population estimate and the number of 9-year-olds obtained from the real GDP observations according to equation (3). Notice the estimated curve behaviour near 1990 - the observed oscillation is considerably smoothed out in the following estimates and in the estimates of 9-year-olds. Initial value of the 9-year-olds in 1951 is 3900000. The actual number of 9-year-olds is overestimated by the projection for the period before 1985 due to a positive demographic dynamics in the USA.

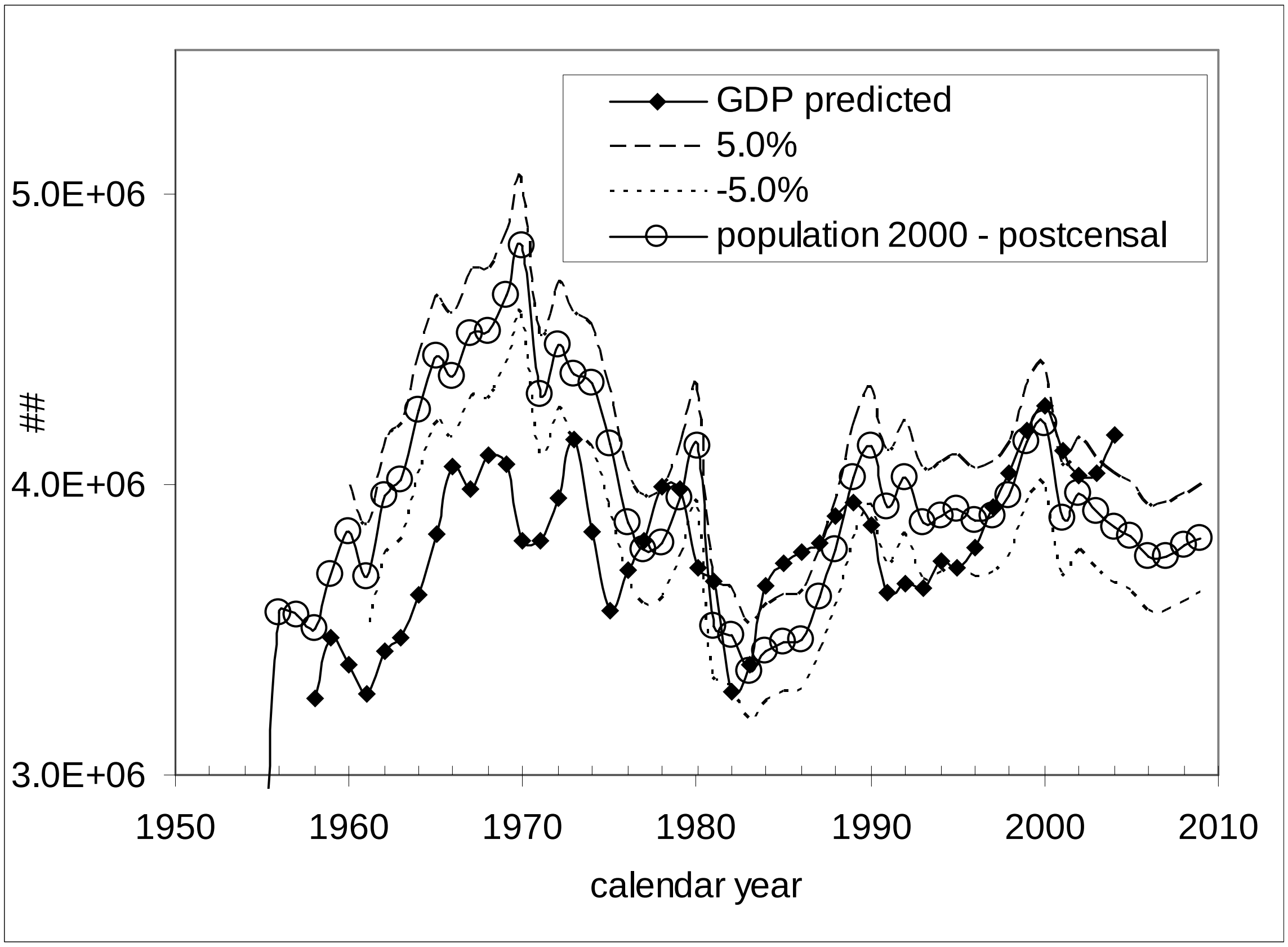

Fig. 13. Comparison of the 9-year-old population as obtained by a back projection from the 2000 postcensal population estimate and the number of 9-year-olds obtained from the real GDP observations according to equation (3). Notice the coincidence of the curves near 2000. Initial value of the 9-year-olds in 1951 is 3800000. The actual number of 9-year-olds is overestimated by the projection for the period before 1975 but in a good agreement between 1975 and 2001.

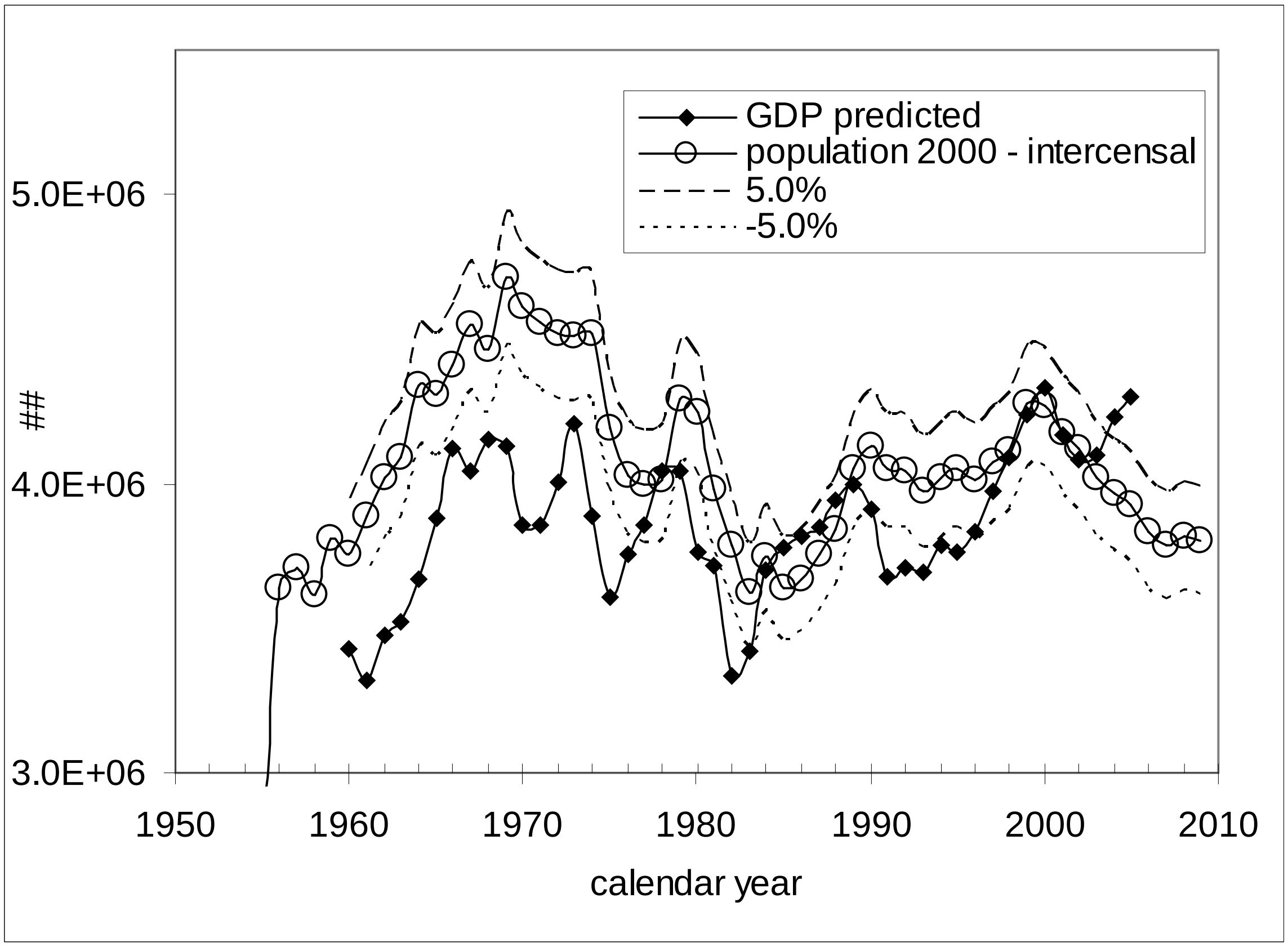

Fig. 14. Comparison of the 9-year-old population as obtained by a back projection from the 2000 intercensal population estimate and the number of 9-year-olds obtained from the real GDP observations according to equation (3). Notice the coincidence of the curves near 2000. Initial value of the 9-year-olds in 1951 is 3800000. The actual number of 9-year-olds is overestimated by the projection for the period before 1975 but in a good agreement between 1975 and 2001.

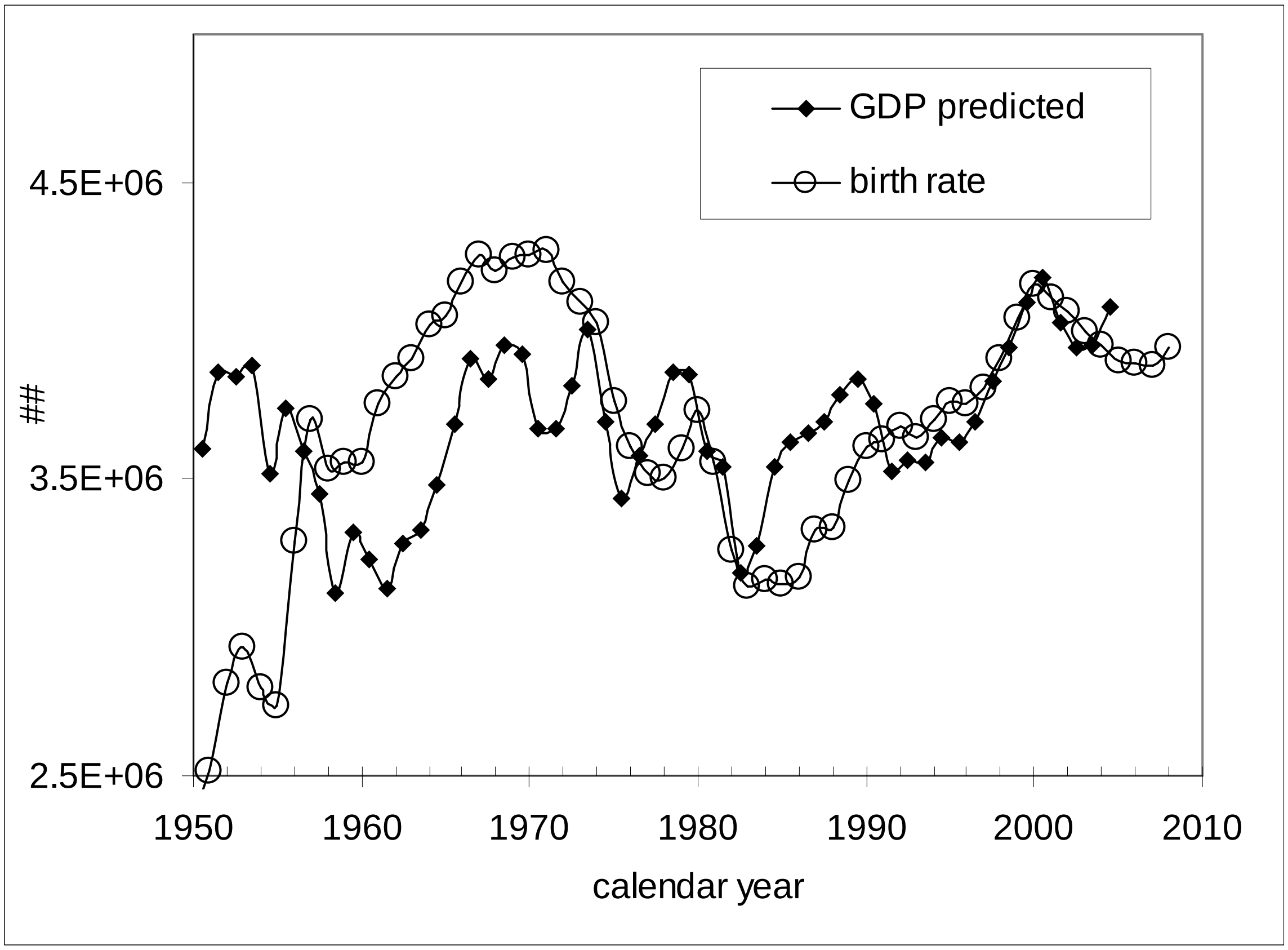

Fig. 15. Comparison of the 9-year-old population as obtained by a back projection from the birth rate (shifted by 9 years ahead) and the number of 9-year-olds obtained from the real GDP observations according to equation (3). Initial value of the 9-year-olds in 1951 is 3860000. The actual number of 9-year-olds is slightly overestimated by the projection for the period before 1970 but in a good agreement between 1975 and 2001. There is no fluctuation in the number of 9-year-olds near 1990 as observed in the 1990 population estimate.

**a)**

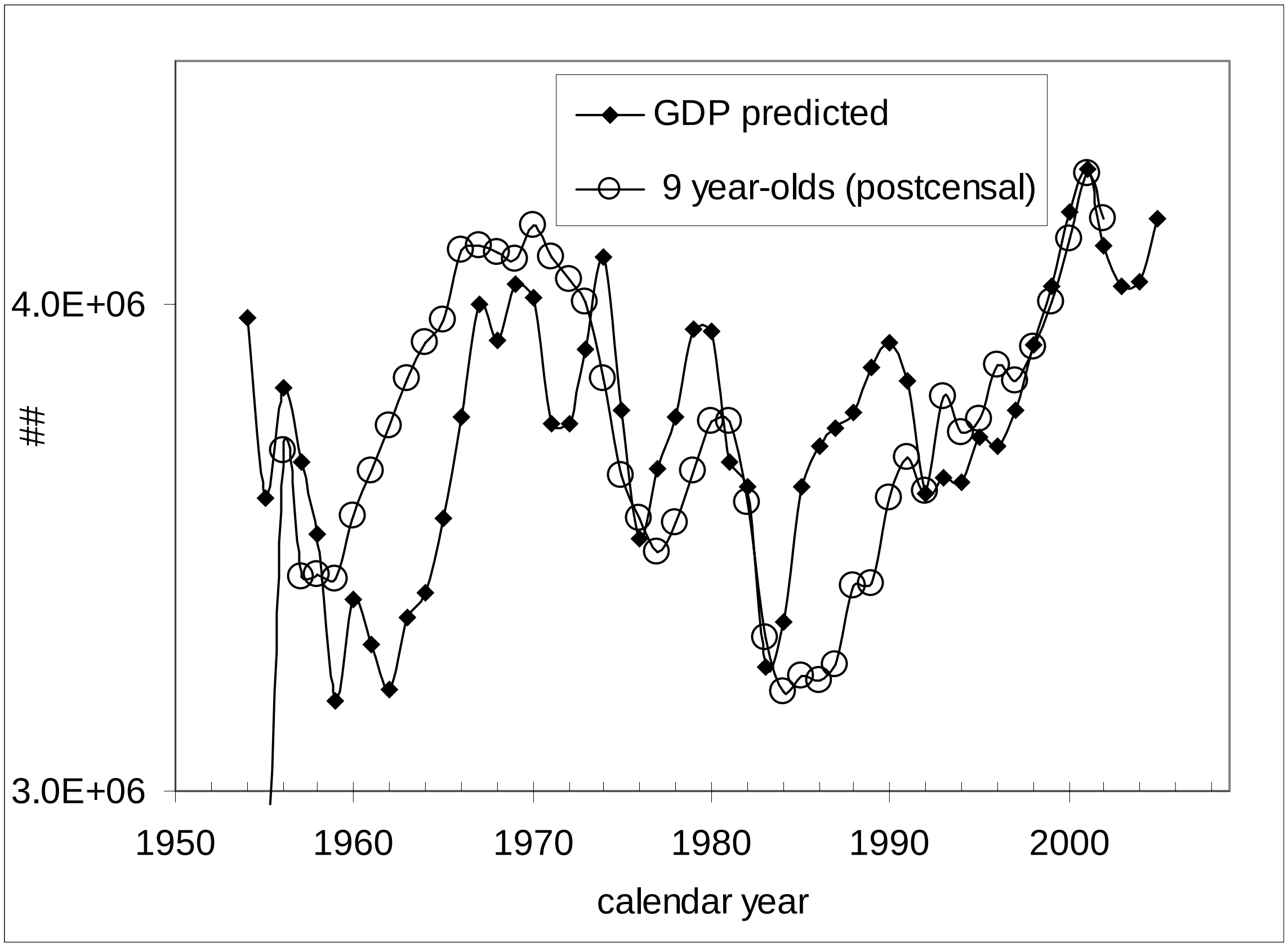

**b)**

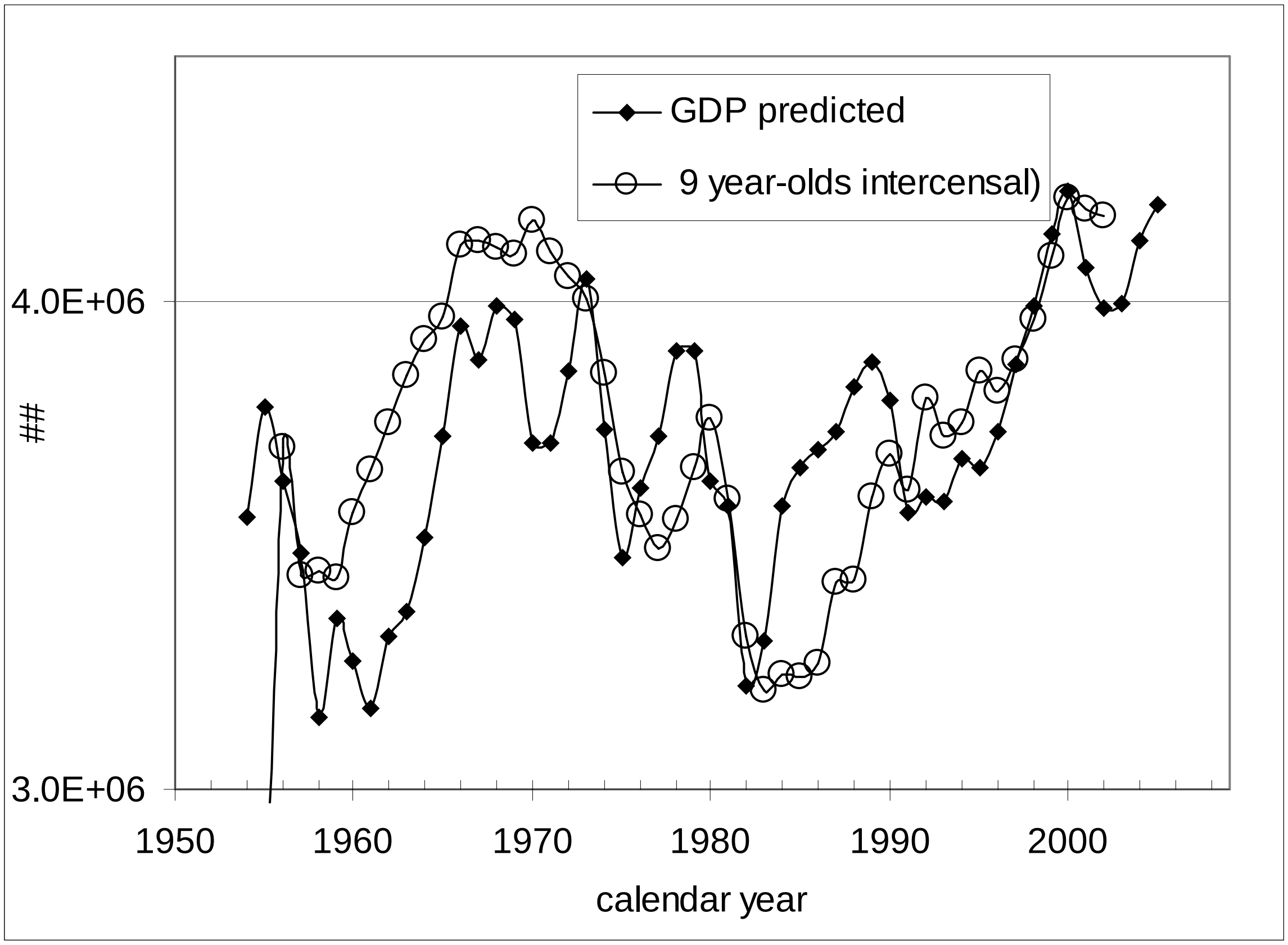

Fig. 16. Comparison of the 9-year-old population - a) the postcensal estimate and b) the intercensal estimate - and the number of 9-year-olds obtained from the real GDP observations according to equation (3). Actual number of 9-year-olds is slightly overestimated by the projection for the period before 1970 but is in a good agreement between 1975 and 2001.

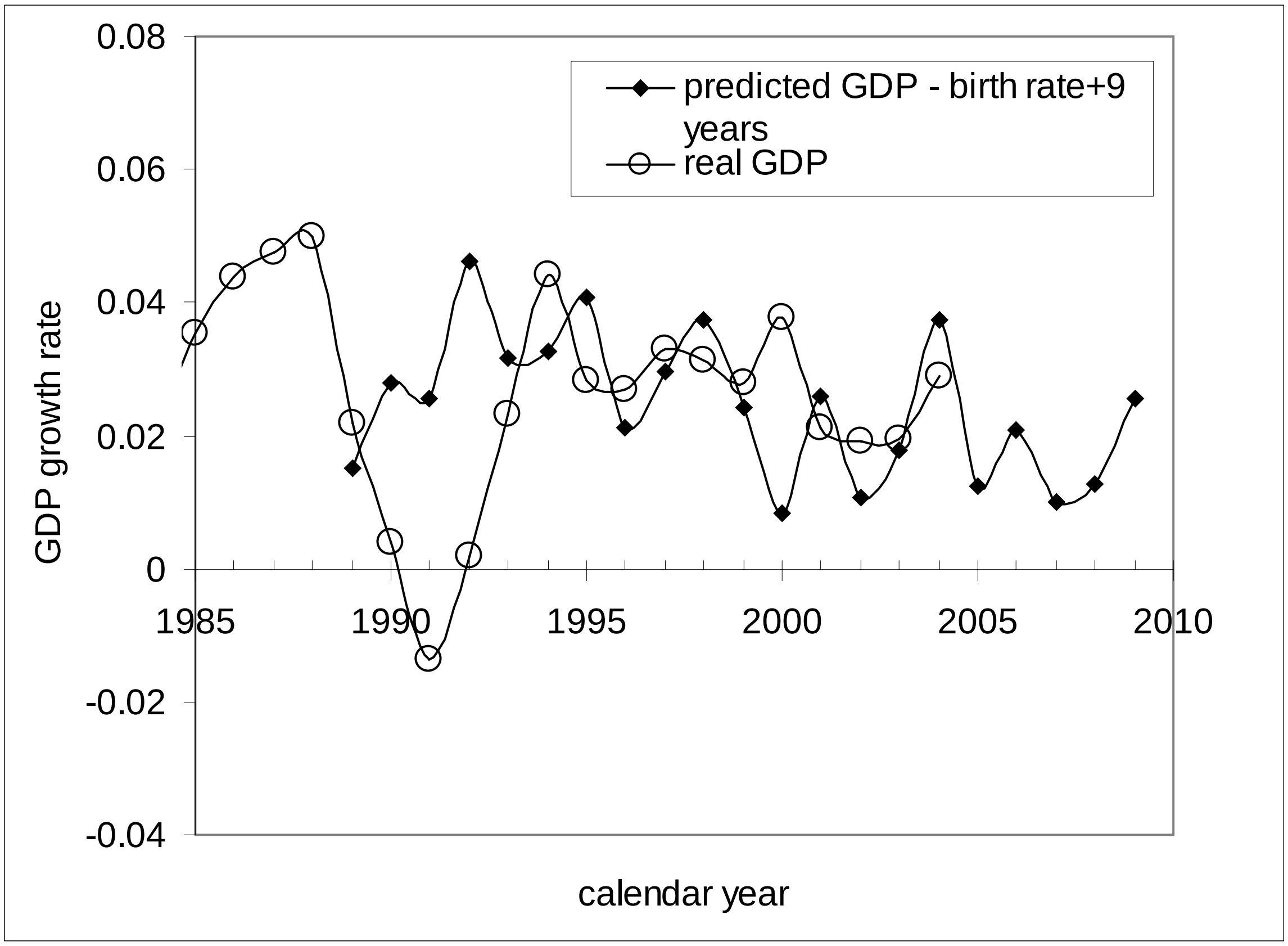

Fig. 17. Real GDP growth rate in the UK for the period between 1985 and 2004. Comparison of the measured values and those predicted by single year of age population estimates (birth rate + 9 years of life):

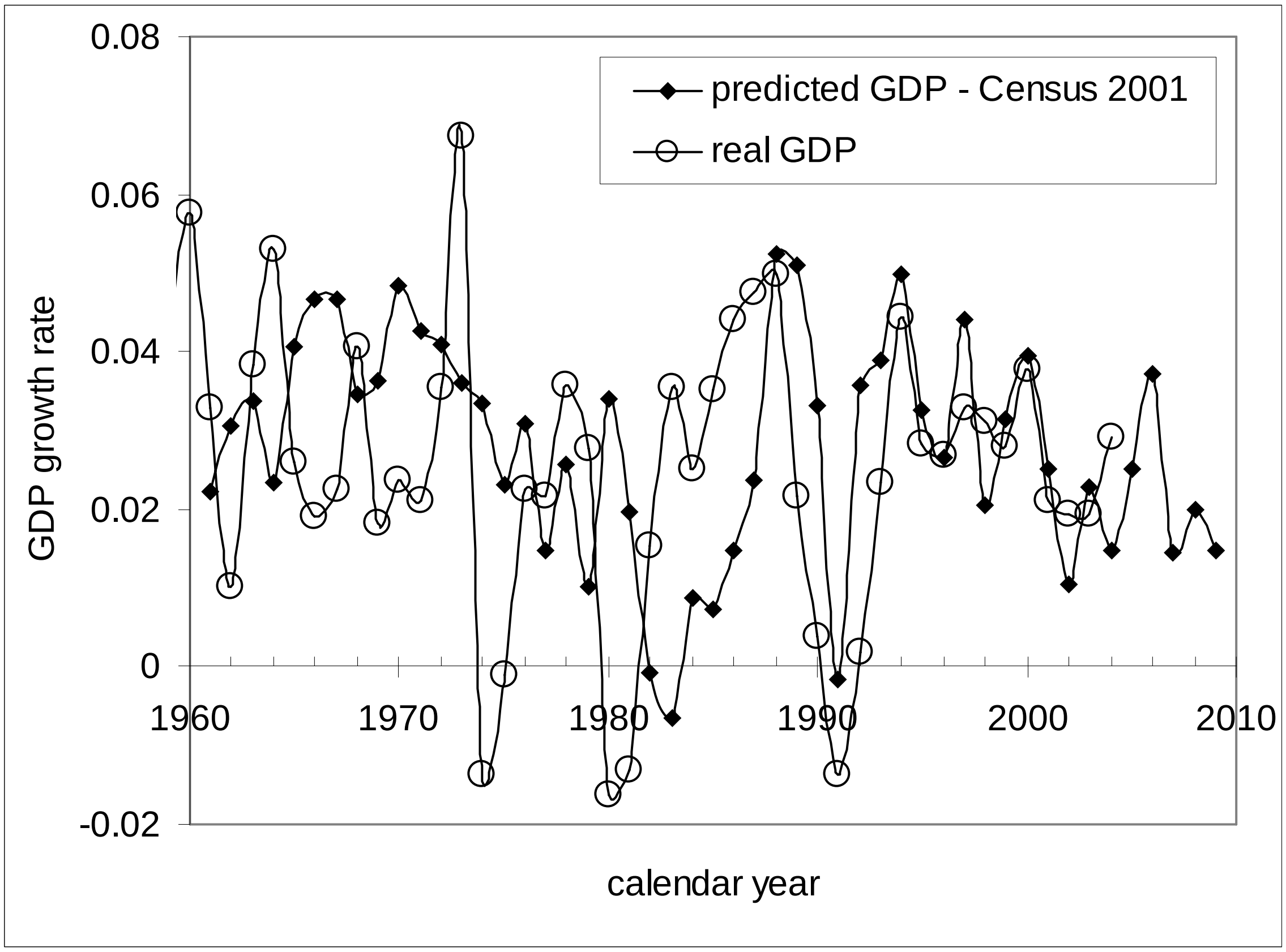

Fig. 18. Real GDP growth rate in the UK for the period between 1960 and 2004. Comparison of the measured values and those predicted by single year of age population estimates (back projection of the 2001 census).

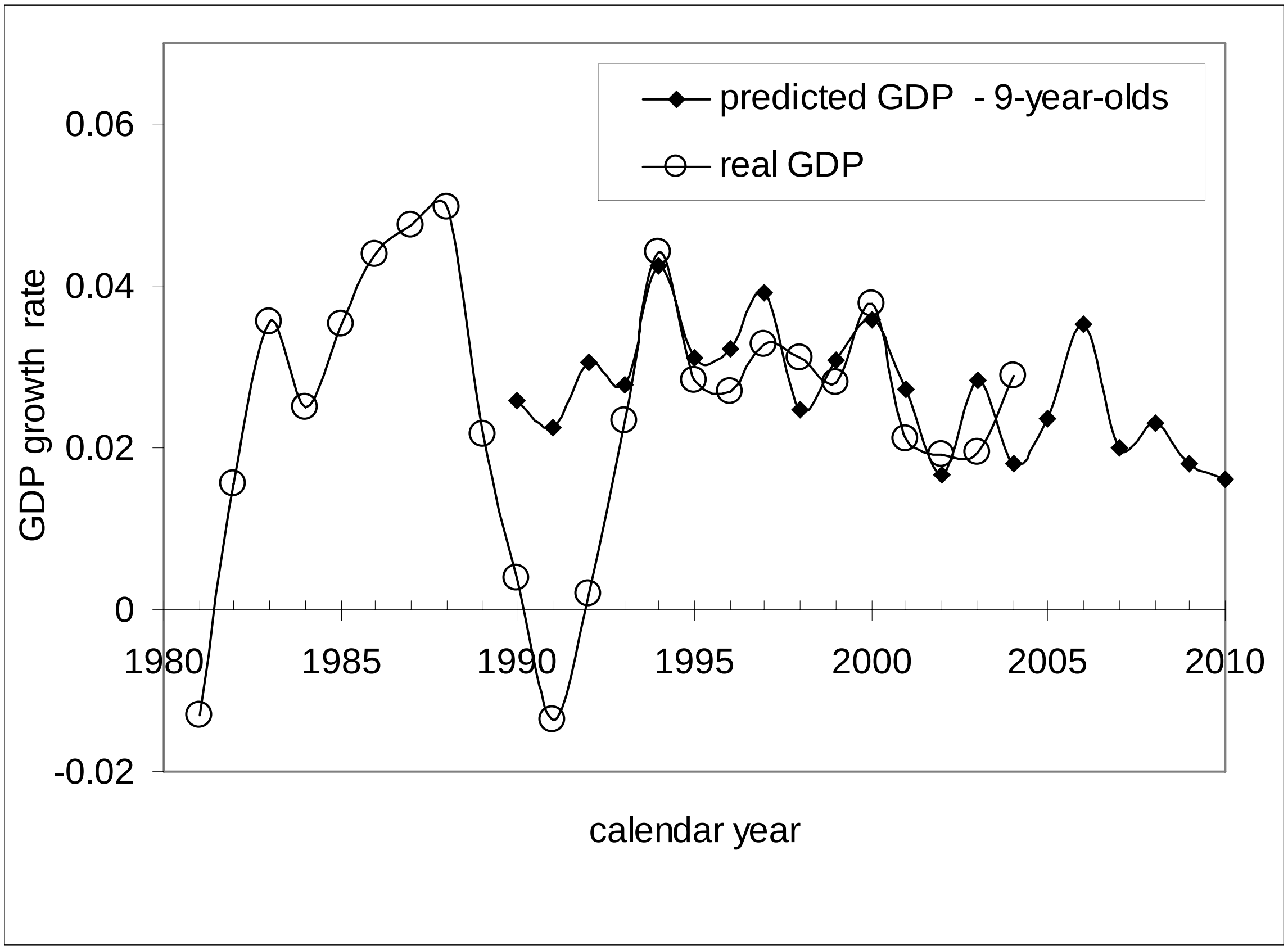

Fig. 19. Real GDP growth rate in the UK for the period between 1990 and 2004. Comparison of the measured values and those predicted by the 9-year-olds estimates.

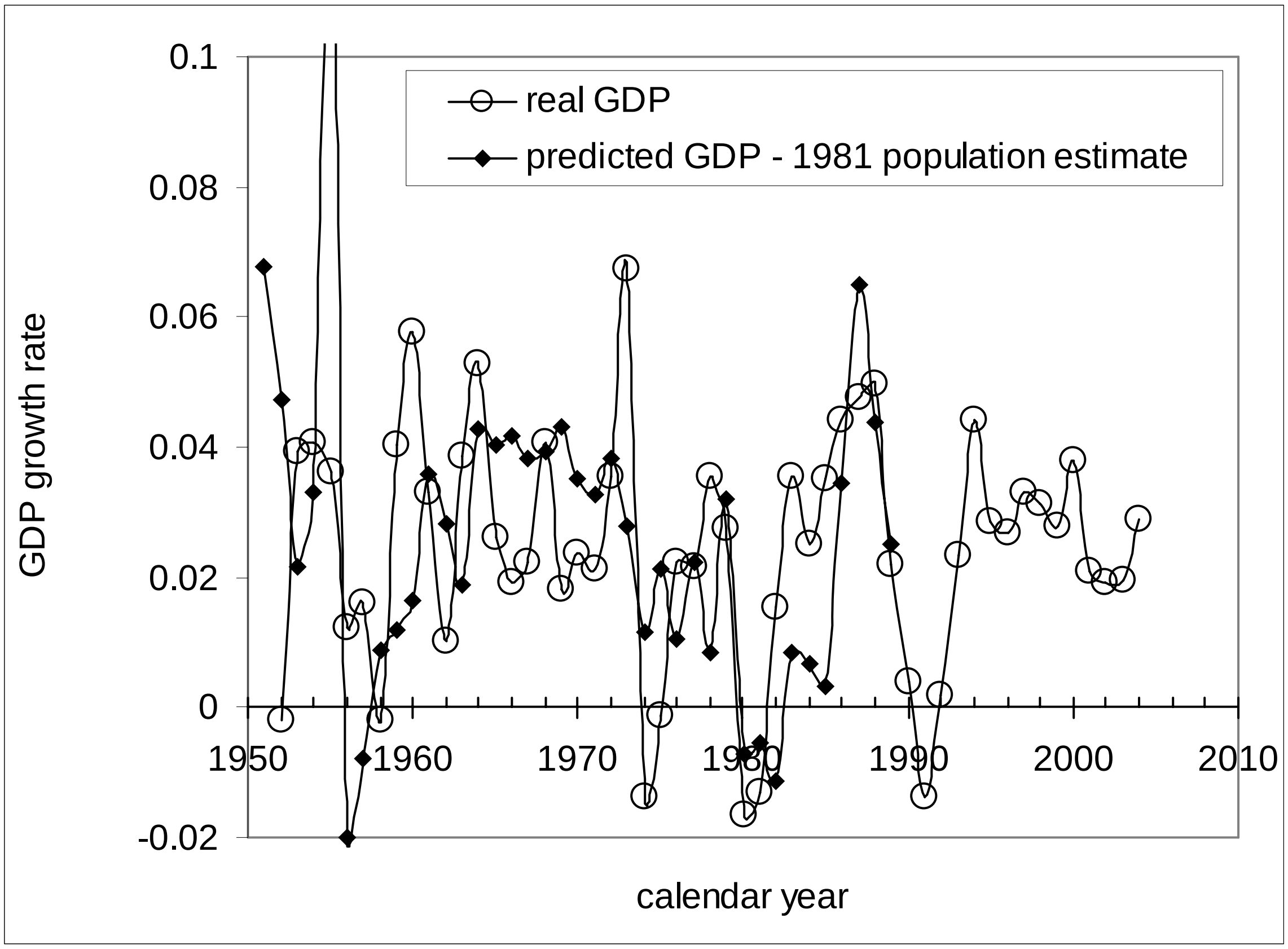

Fig. 20. Real GDP growth rate in the UK for the period between 1950 and 2004. Comparison of the measured values and values predicted by single year of age population estimates (back projection of the 1981 population estimate). Notice the years around 1980.

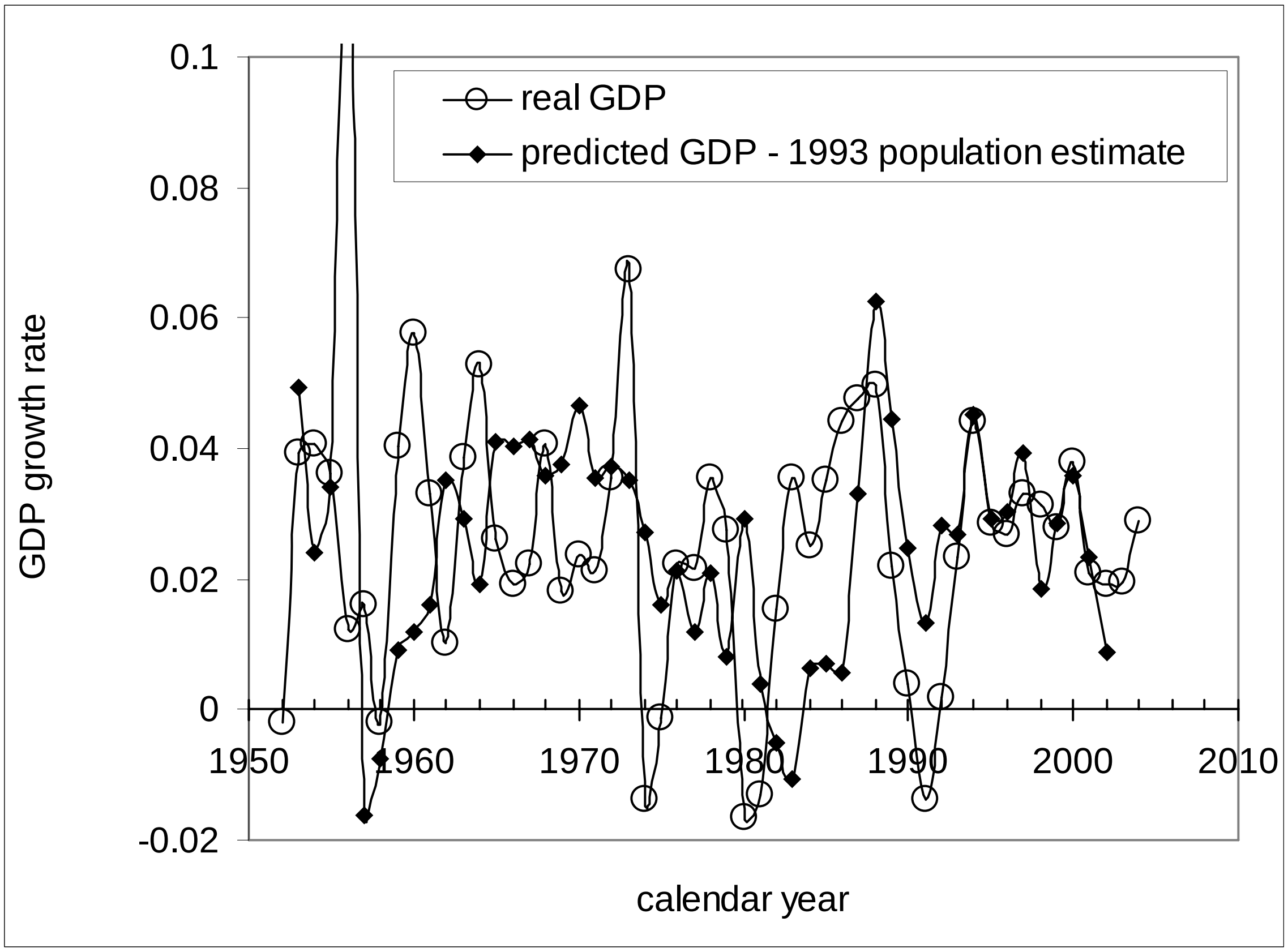

Fig. 21. Real GDP growth rate in the UK for the period between 1950 and 2004. Comparison of the measured values and values predicted by single year of age population estimates (back projection of the 1993 population estimate). Notice the years around 1990.

**a)**

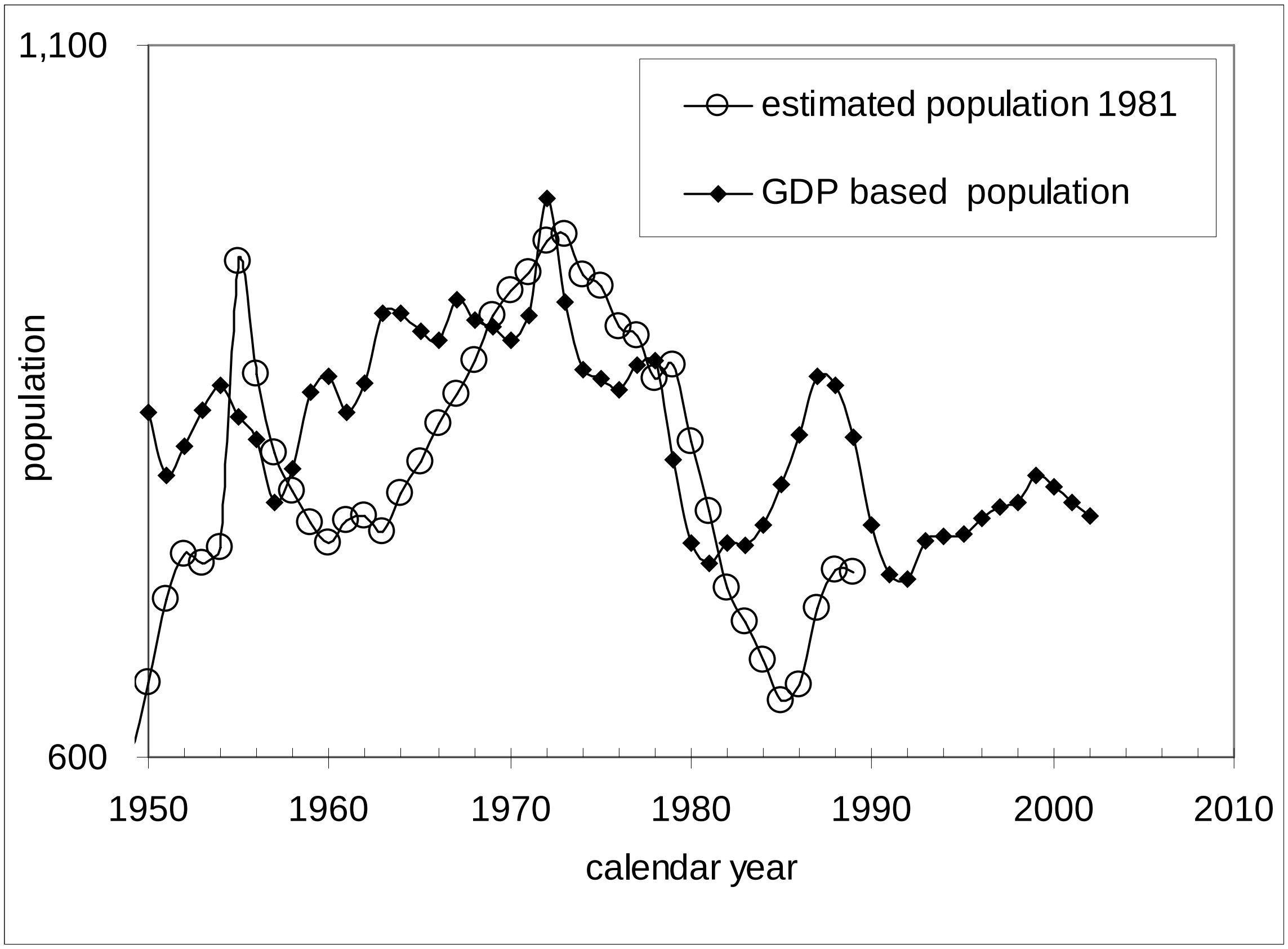

**b)**

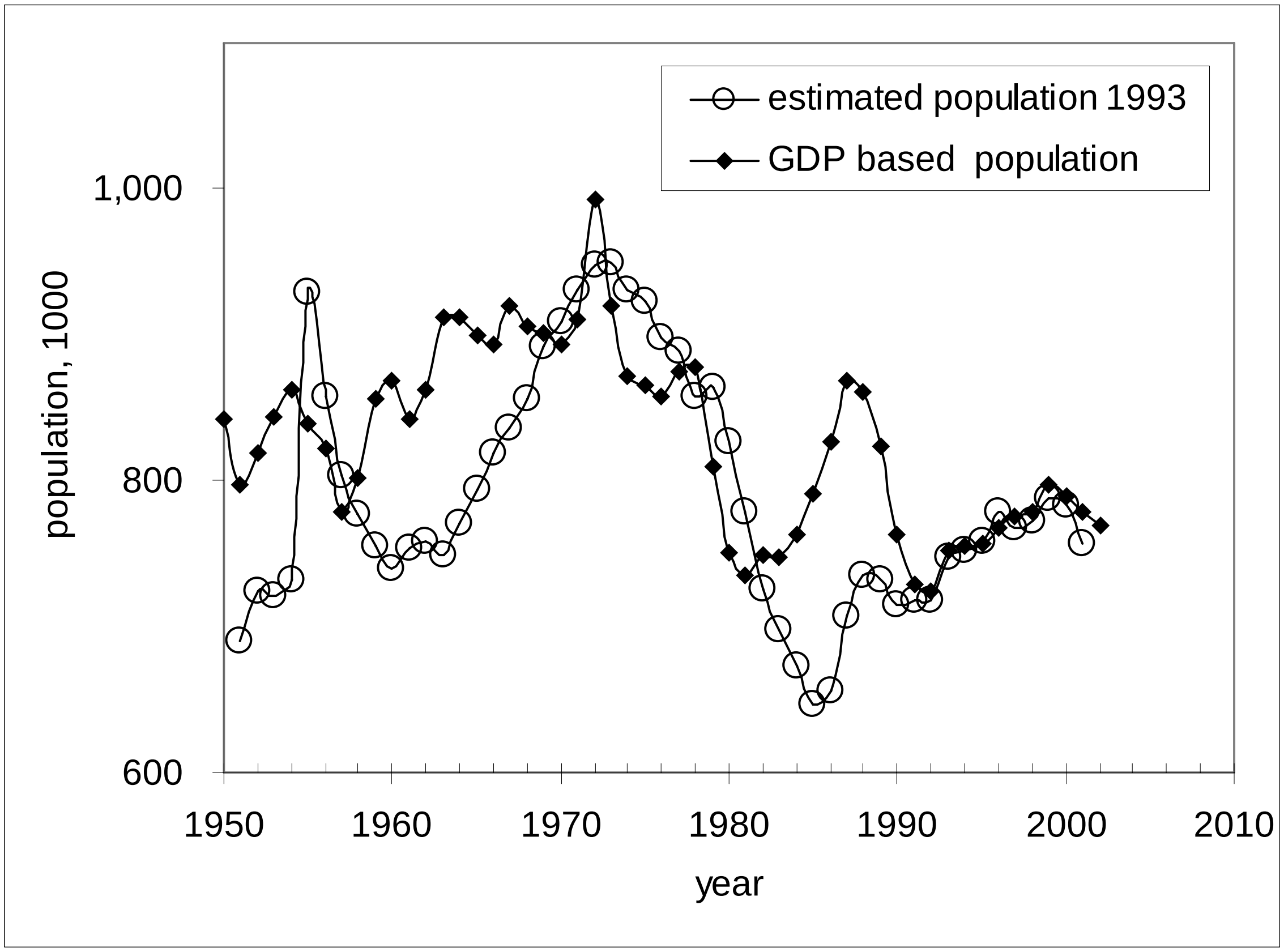

**c)**

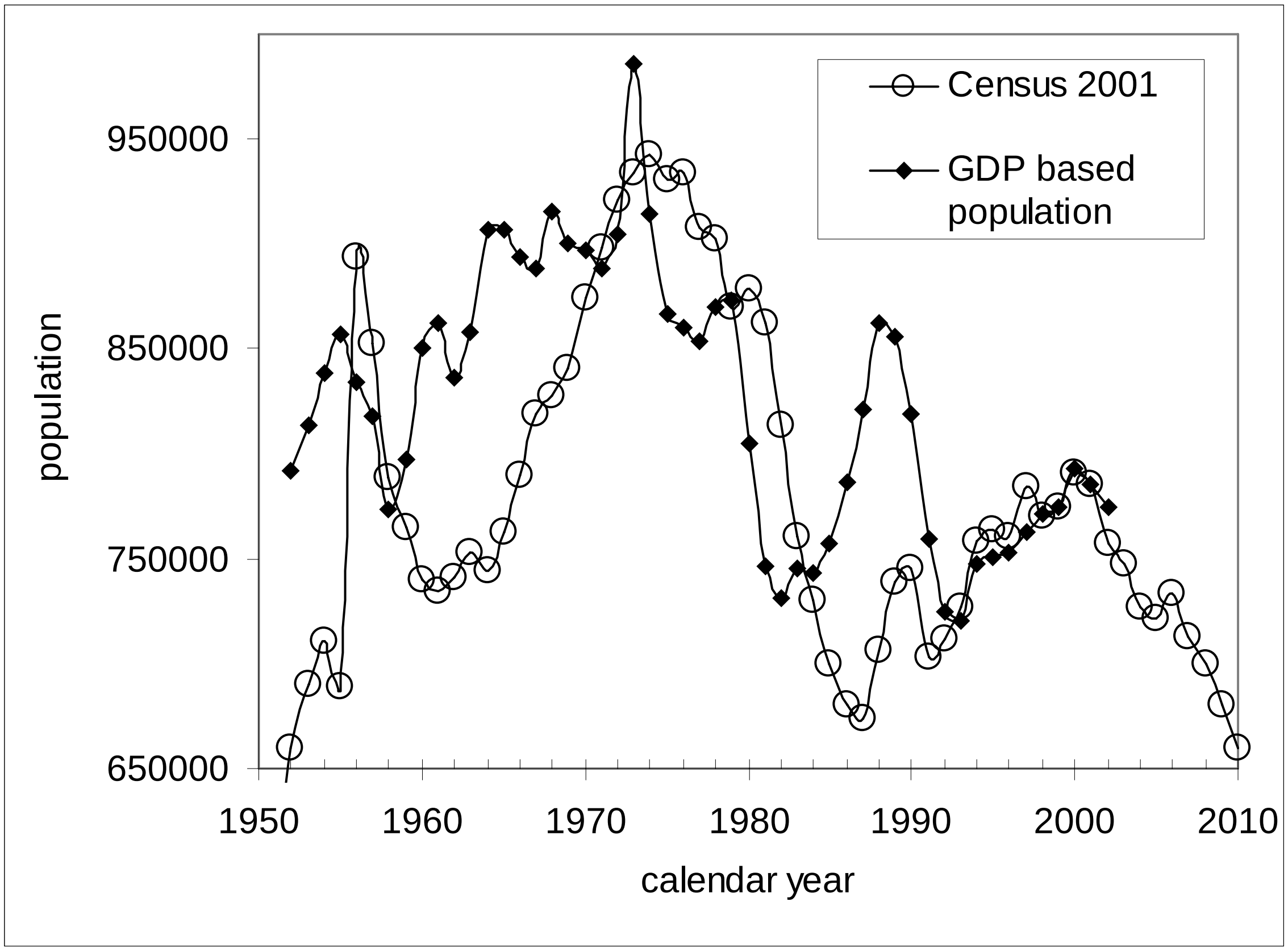

Fig. 22. Comparison of the 9-year-old population in the UK as obtained by a back projection from: **a)** the 1981 population estimate; **b)** the 1993 population estimate; **c)** the 2001 census and the number of 9-year-olds obtained from the real GDP observations according to equation (3).

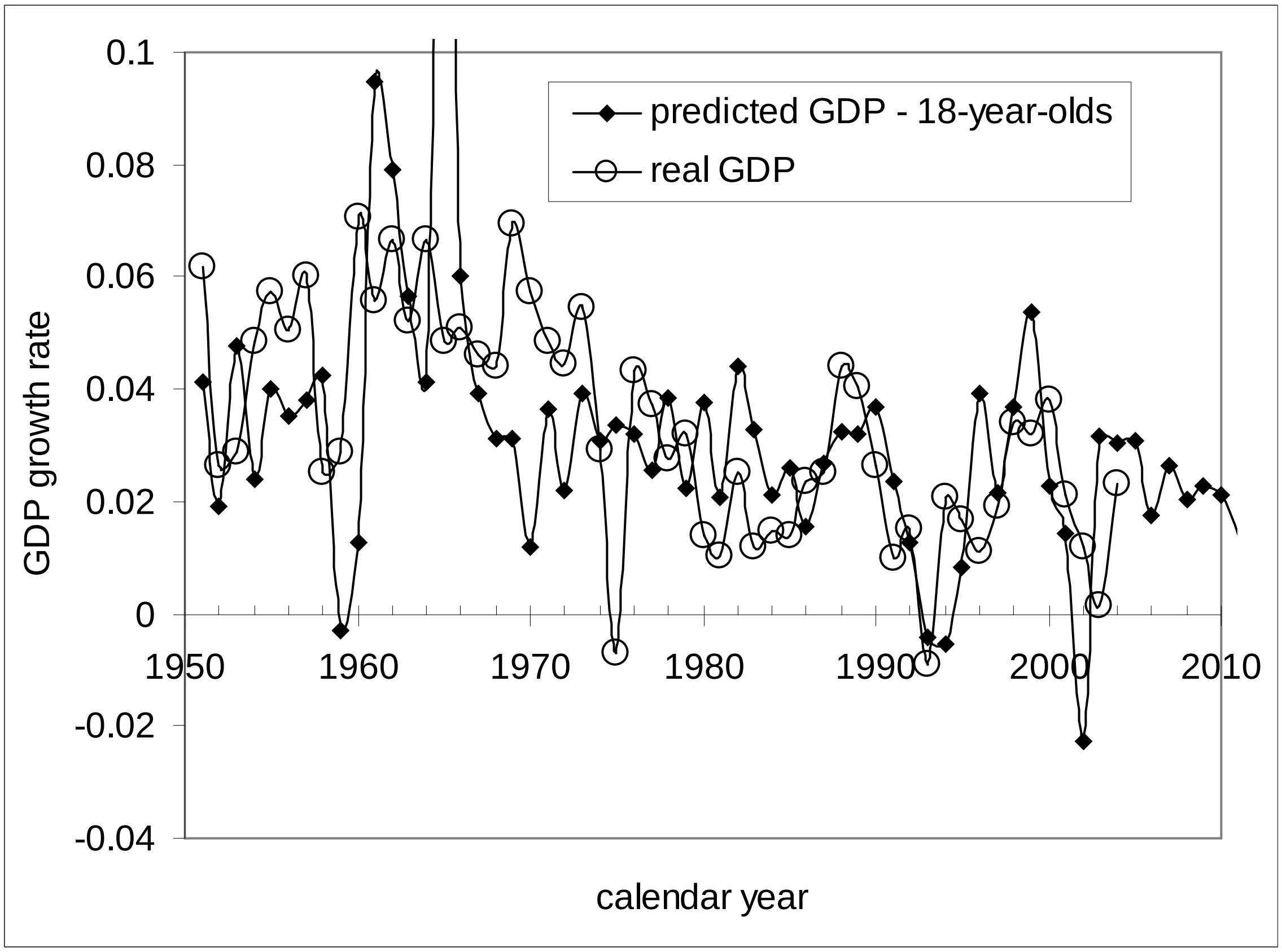

Fig. 23. Real GDP growth rate in France for the period between 1950 and 2004. Comparison of the measured values and values predicted by single year of age (18-year-olds) population estimates.

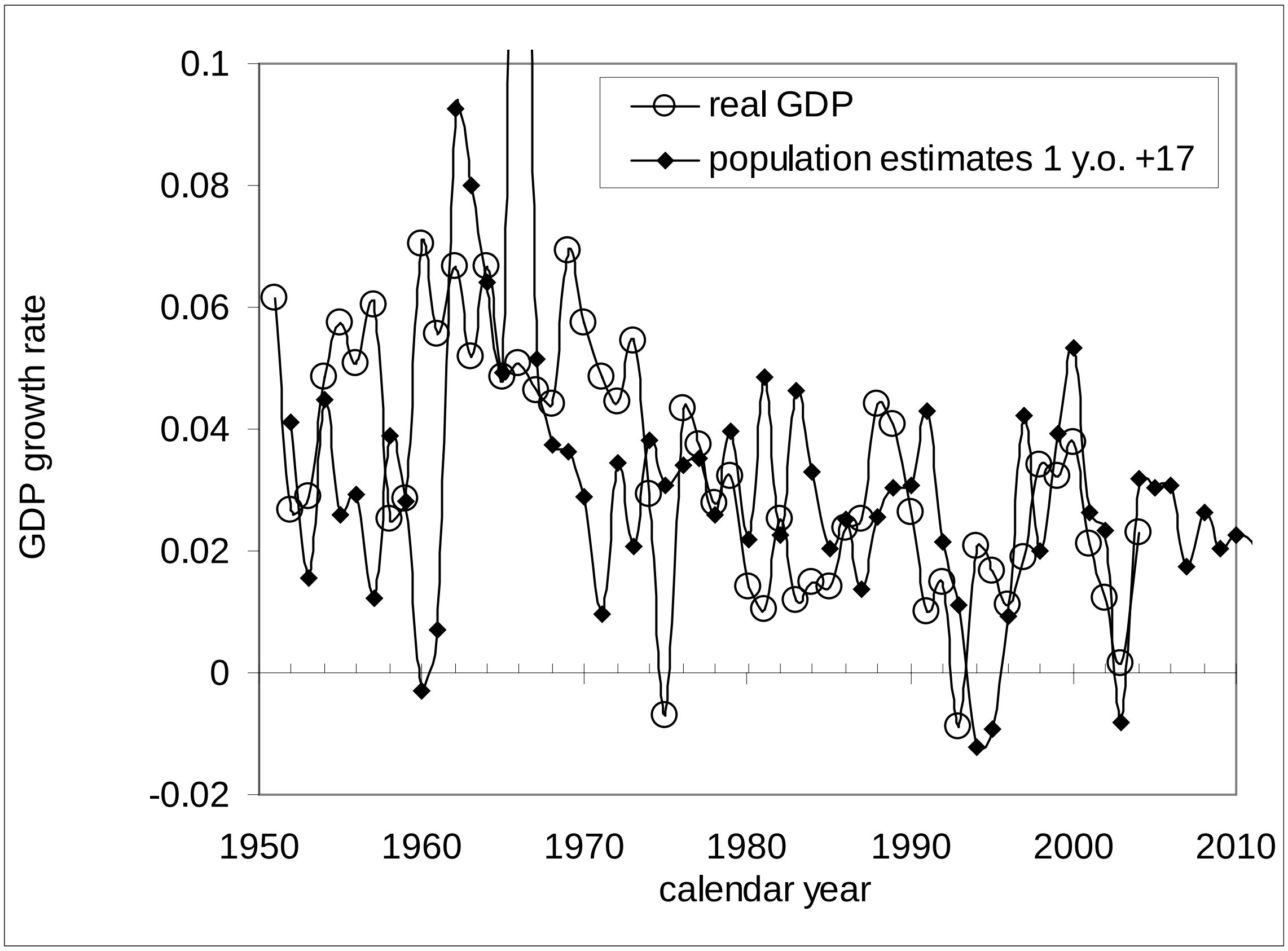

Fig. 24. Real GDP growth rate in France for the period between 1950 and 2004. Comparison of the measured values and values predicted by single year of age (1-year-olds shifted ahead by 17 years) population estimates.

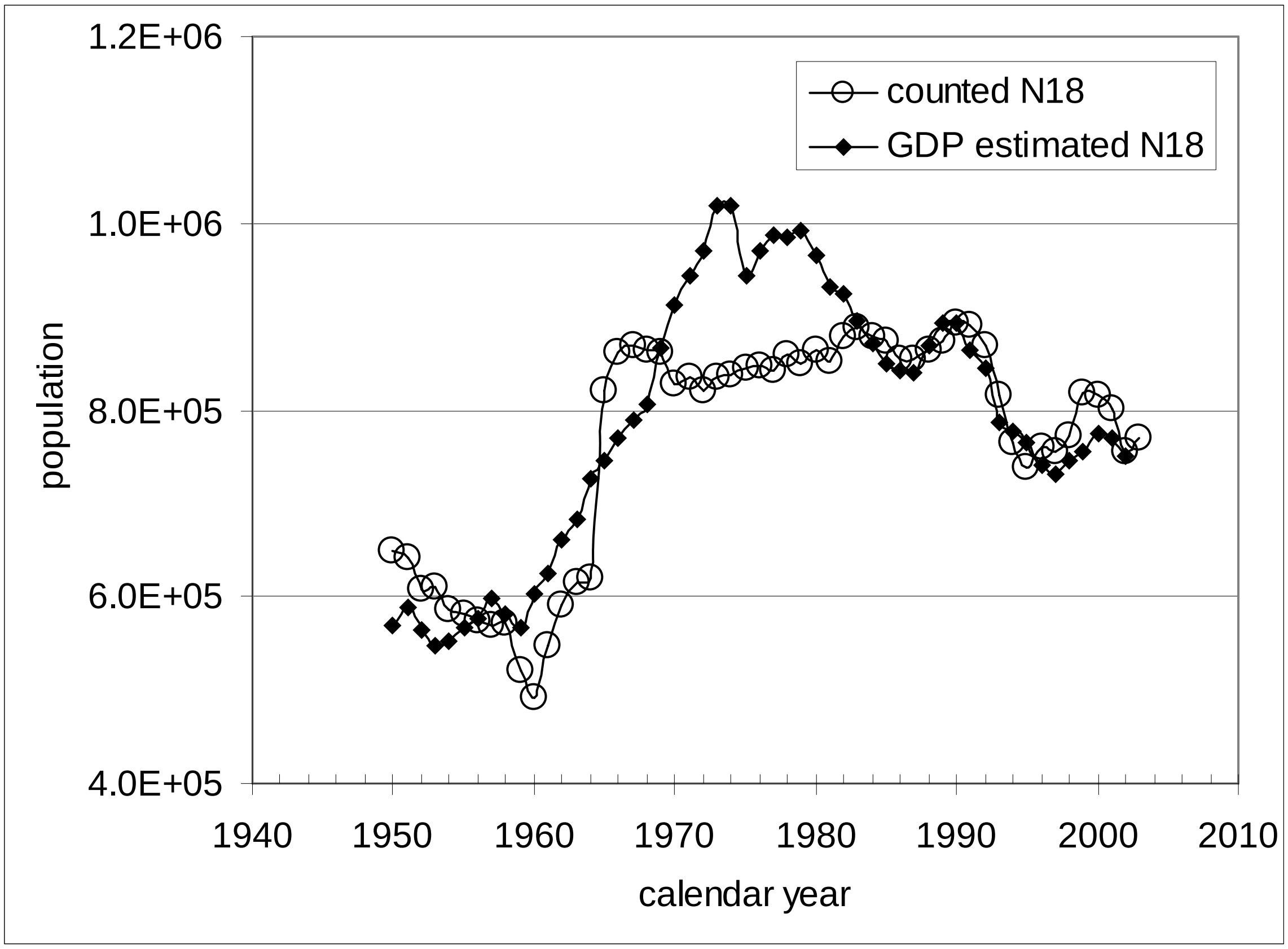

Fig. 25. Comparison of the estimated 18-year-old population in France and the number of 18-year-olds obtained from the real GDP observations according to equation (3).

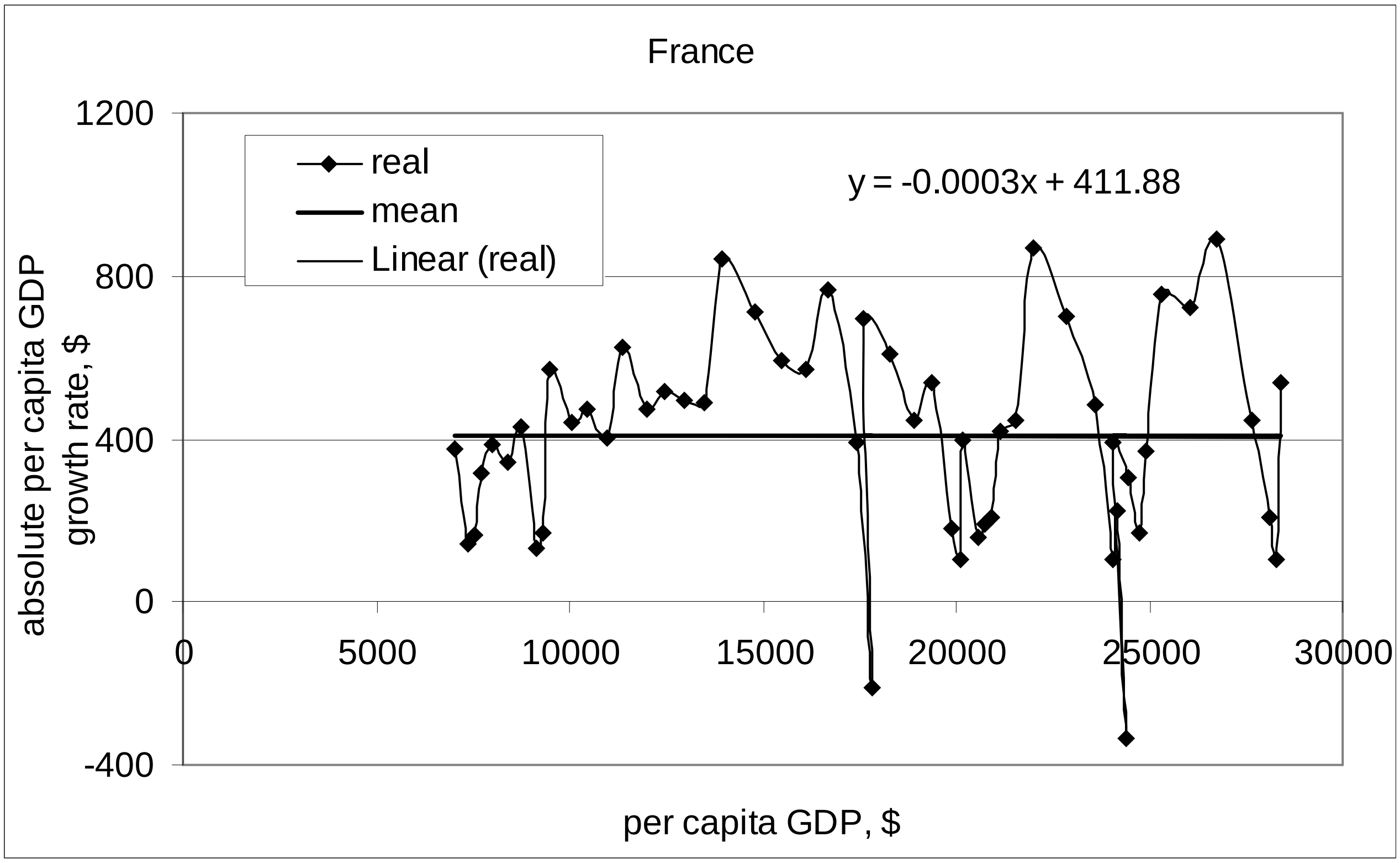

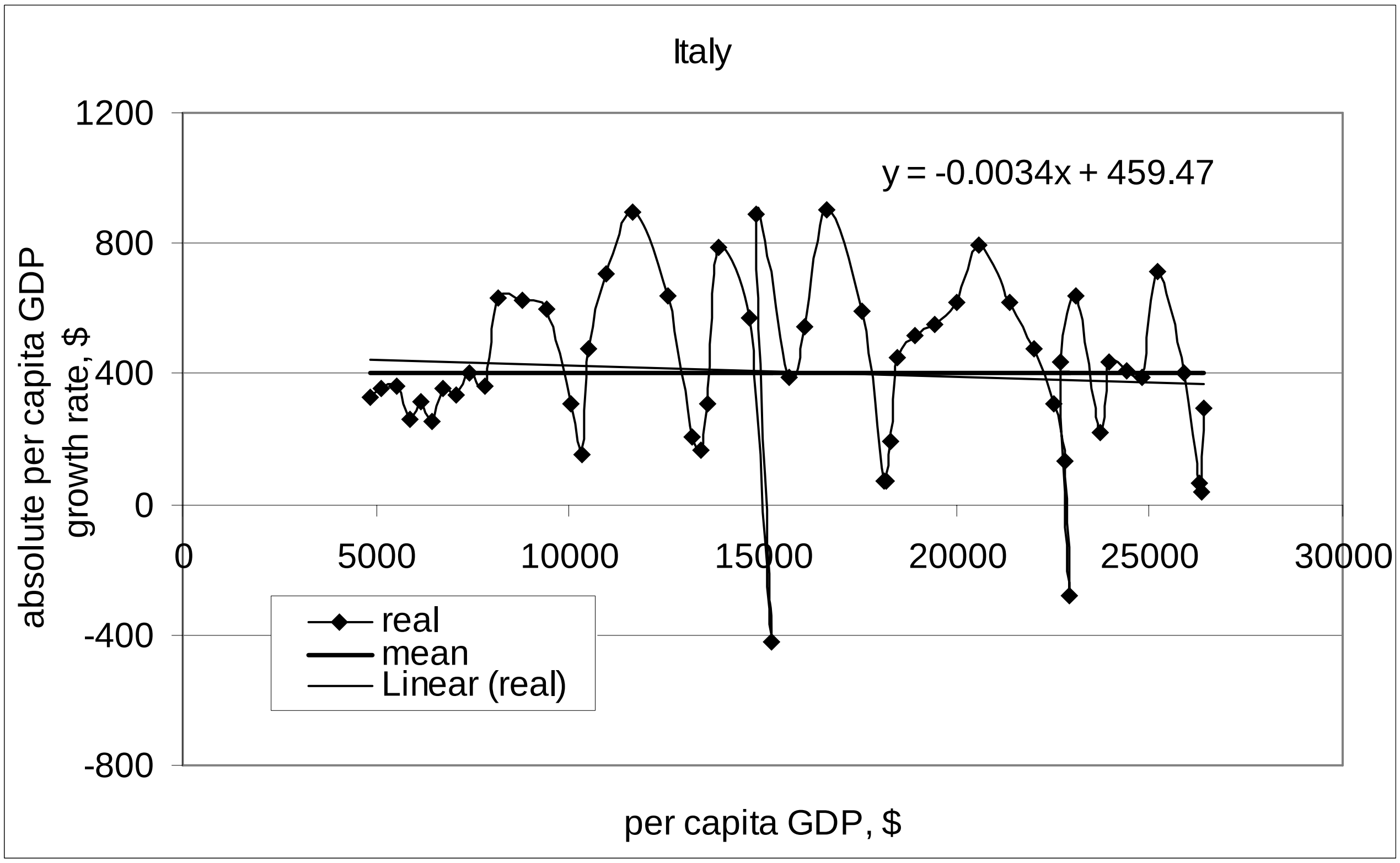

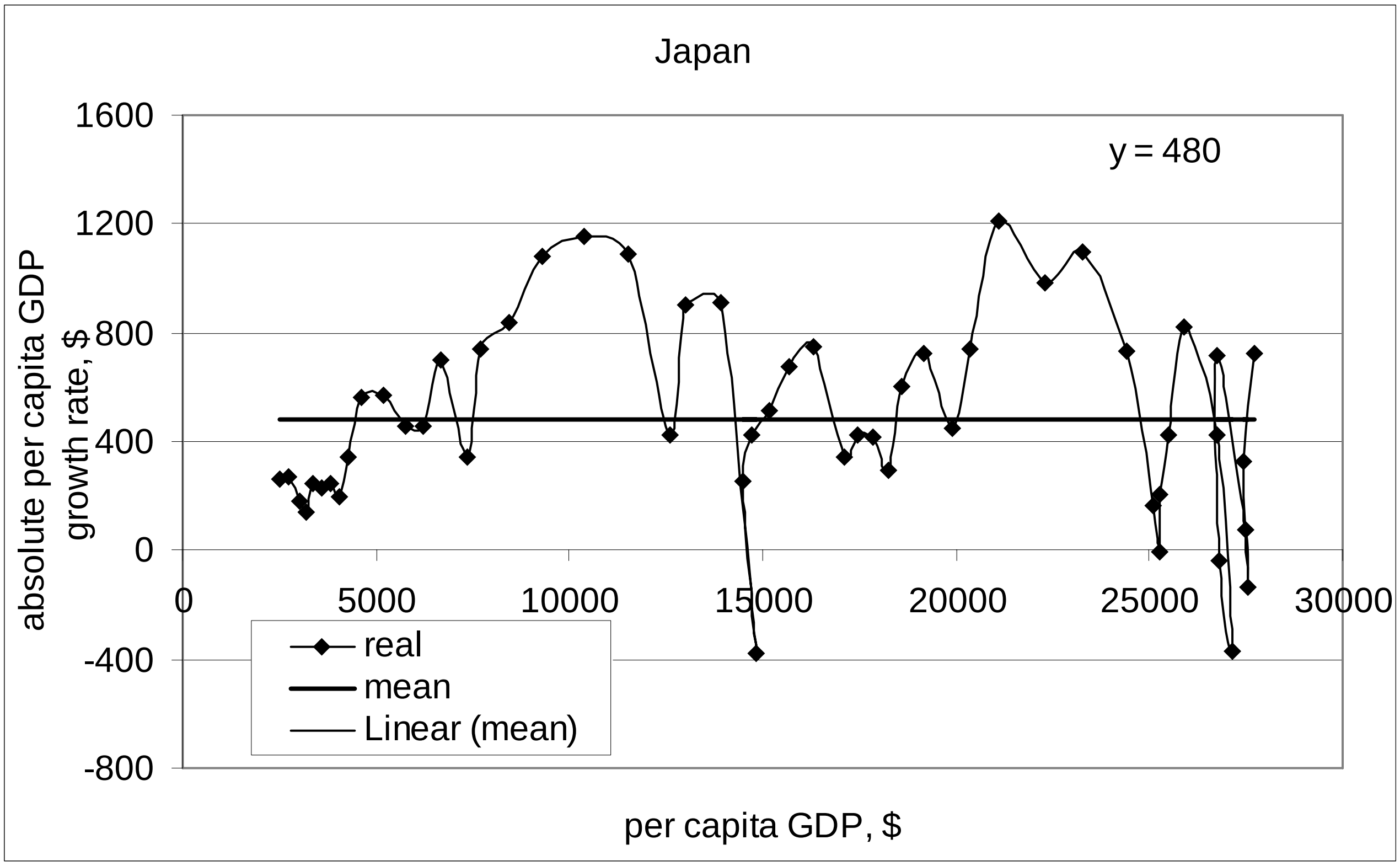

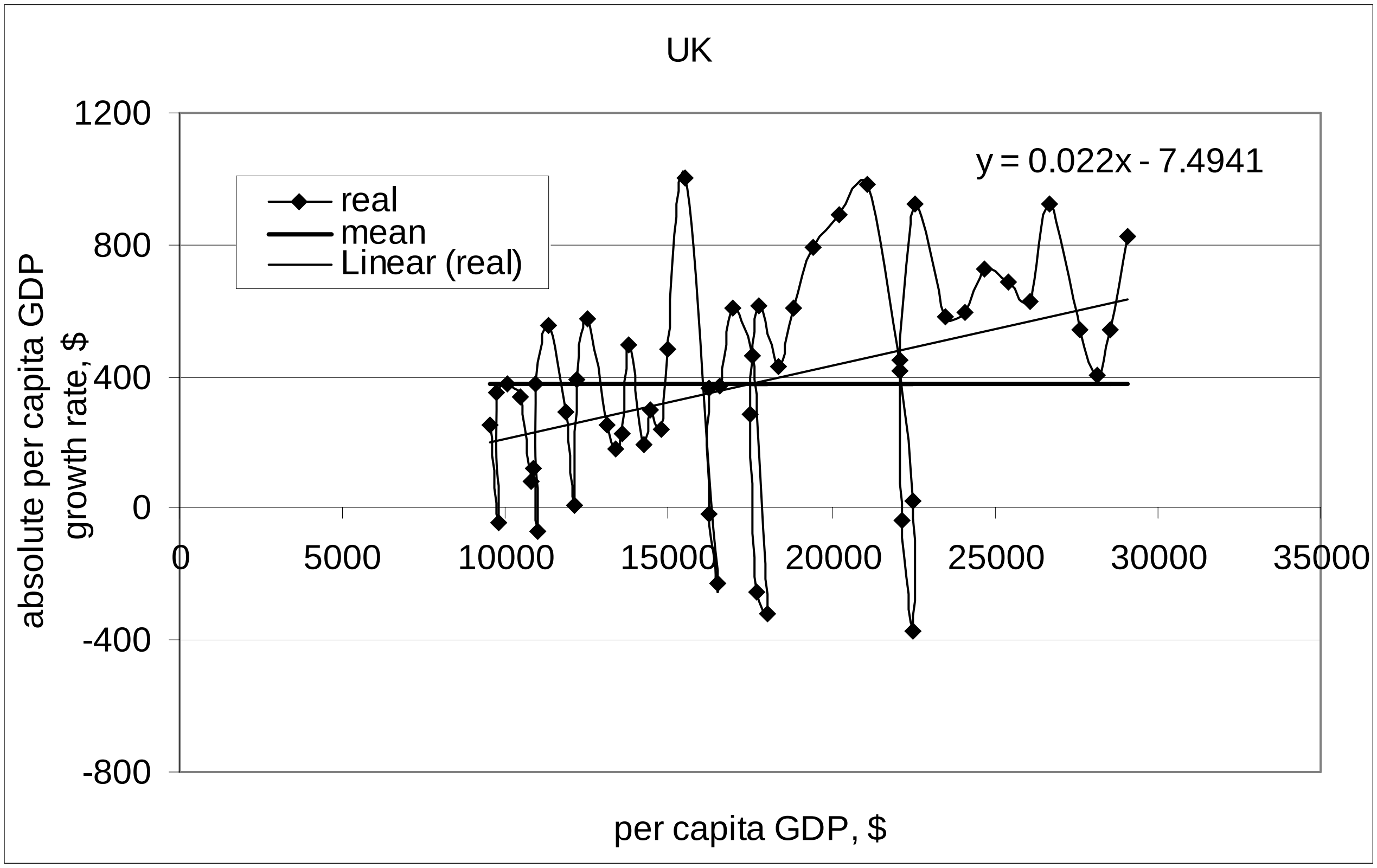

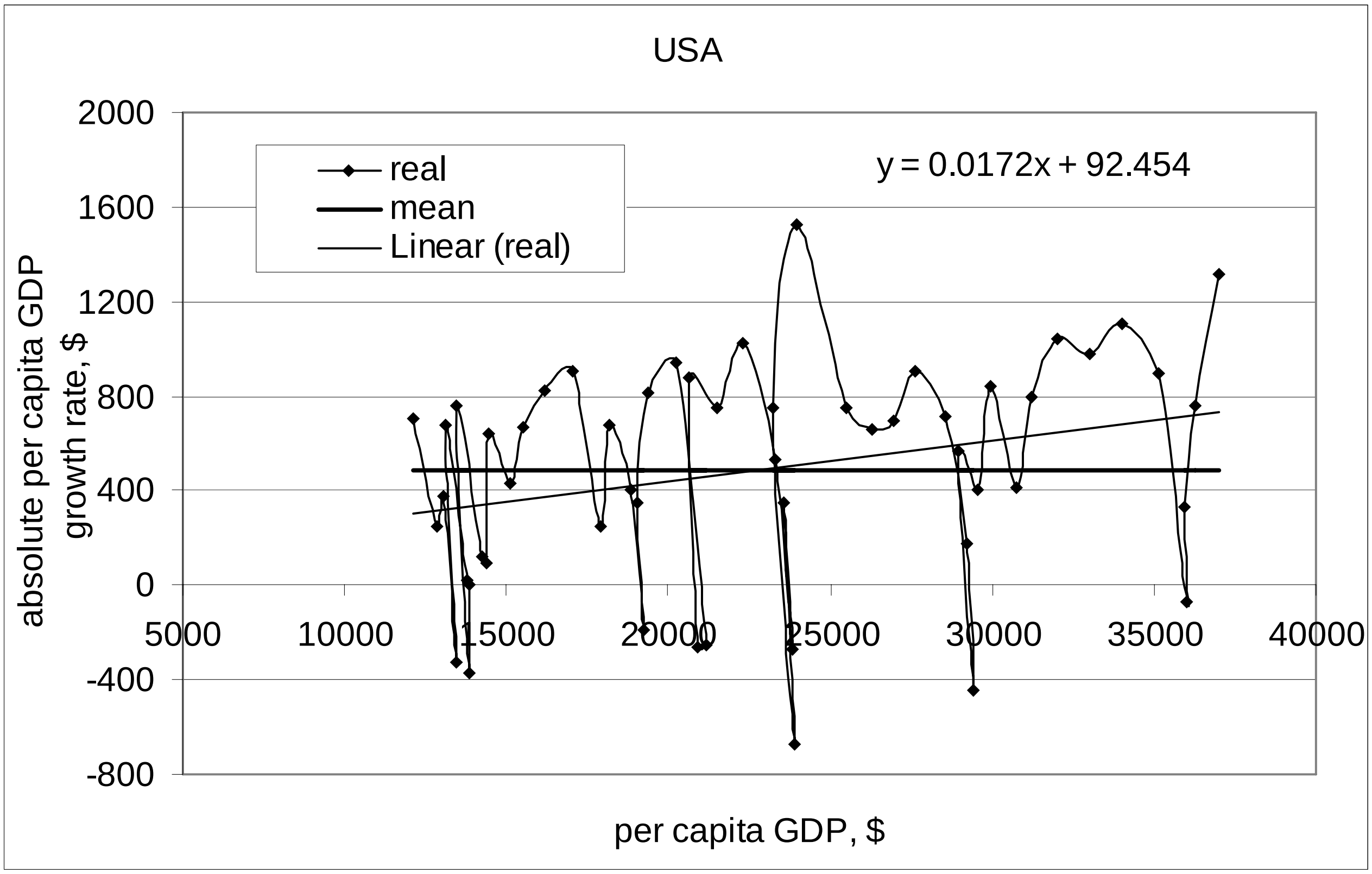

Fig. 26. Approximation of the observed GDP per capita *annual increments* by a constant for France, Italy, Japan, the UK, and the USA. When applied, a linear regression gives the same line for France, Italy, and Japan. For the UK and the USA, the linear regression indicates some increase in the GDP per capita annual increment. This increase is explained by the effect of the population component change.

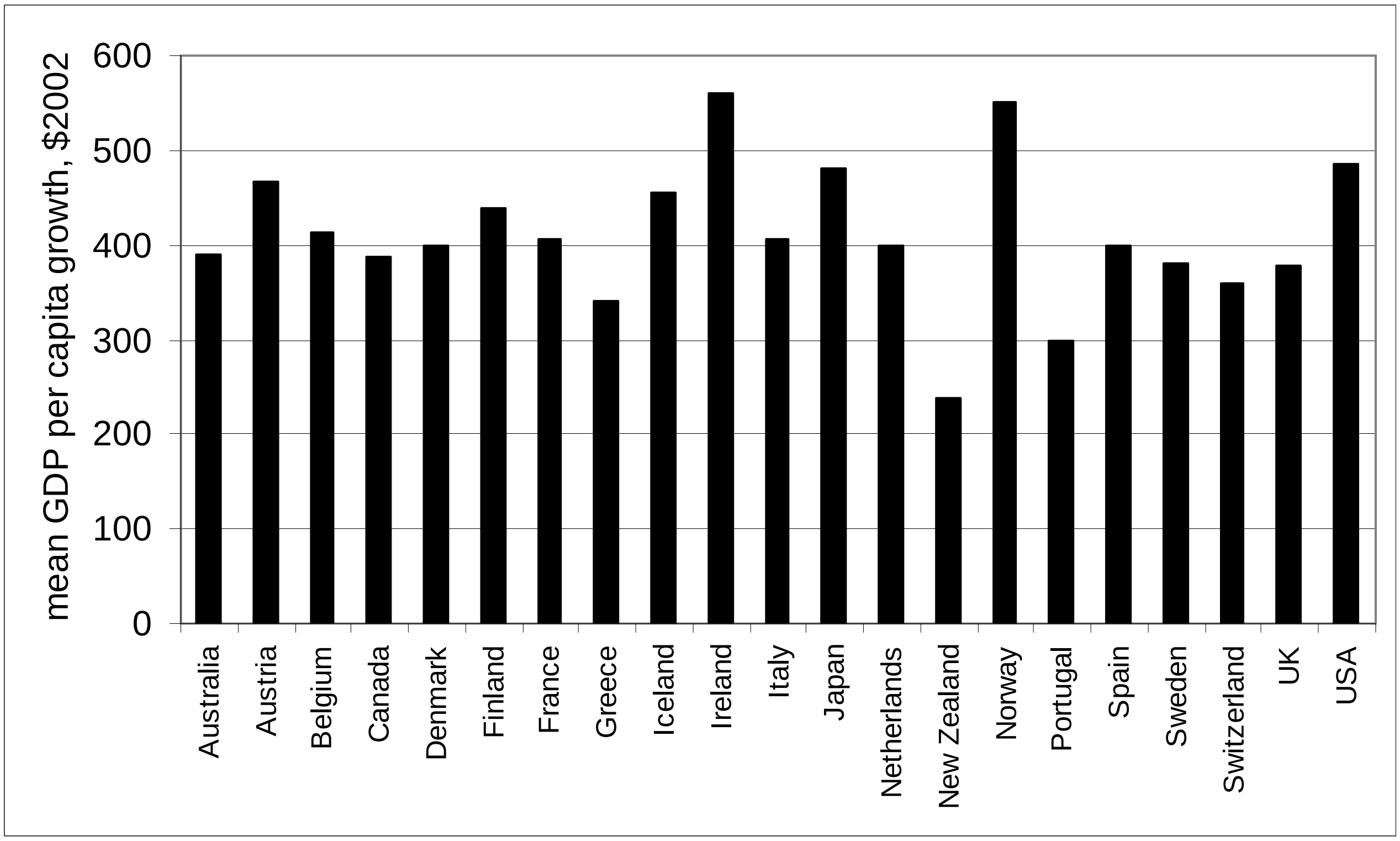

Fig. 27. Mean per capita GDP growth in some developed countries. The values are near $400 per year for most of the countries. Some small economies are characterized by elevated values.  When corrected for the population component of growth, the USA has the same value of about $400 per year. Greece and Portugal have been underperforming for a long time in past and New Zealand as well.

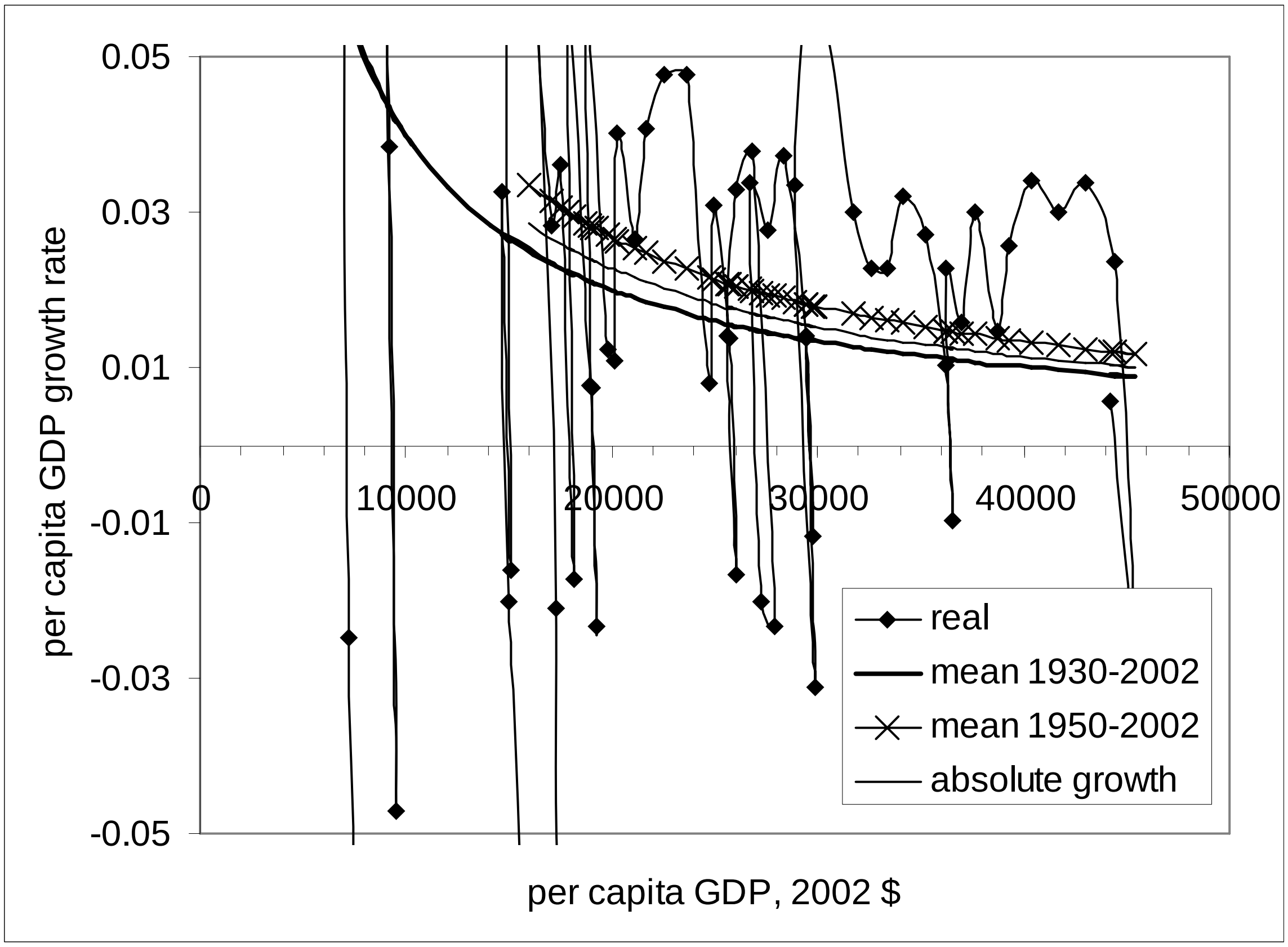

Fig. 28. Approximation of the observed per capita GDP growth rate by a function A/G (A is the absolute value of constant growth, G is the GDP per capita) for the USA. Black and red curves correspond to the mean absolute growth values for the period between 1930 and 2002 and the period between 1950 and 2002, respectively. Magenta curve illustrates the relative growth related to economic trend only, i.e. the total economic growth less the population component. All the per capita GDP values are corrected for the difference between the total population and the population of 15 years old and above (economically active population). The ratio of the populations in the USA changed from of 1.37 in 1950 to 1.27 in 2002. In 1930, the ratio was 1.41. Thus the per capita GDP values are larger by 1.37/1.27 times than in the original table used also for other countries under study.

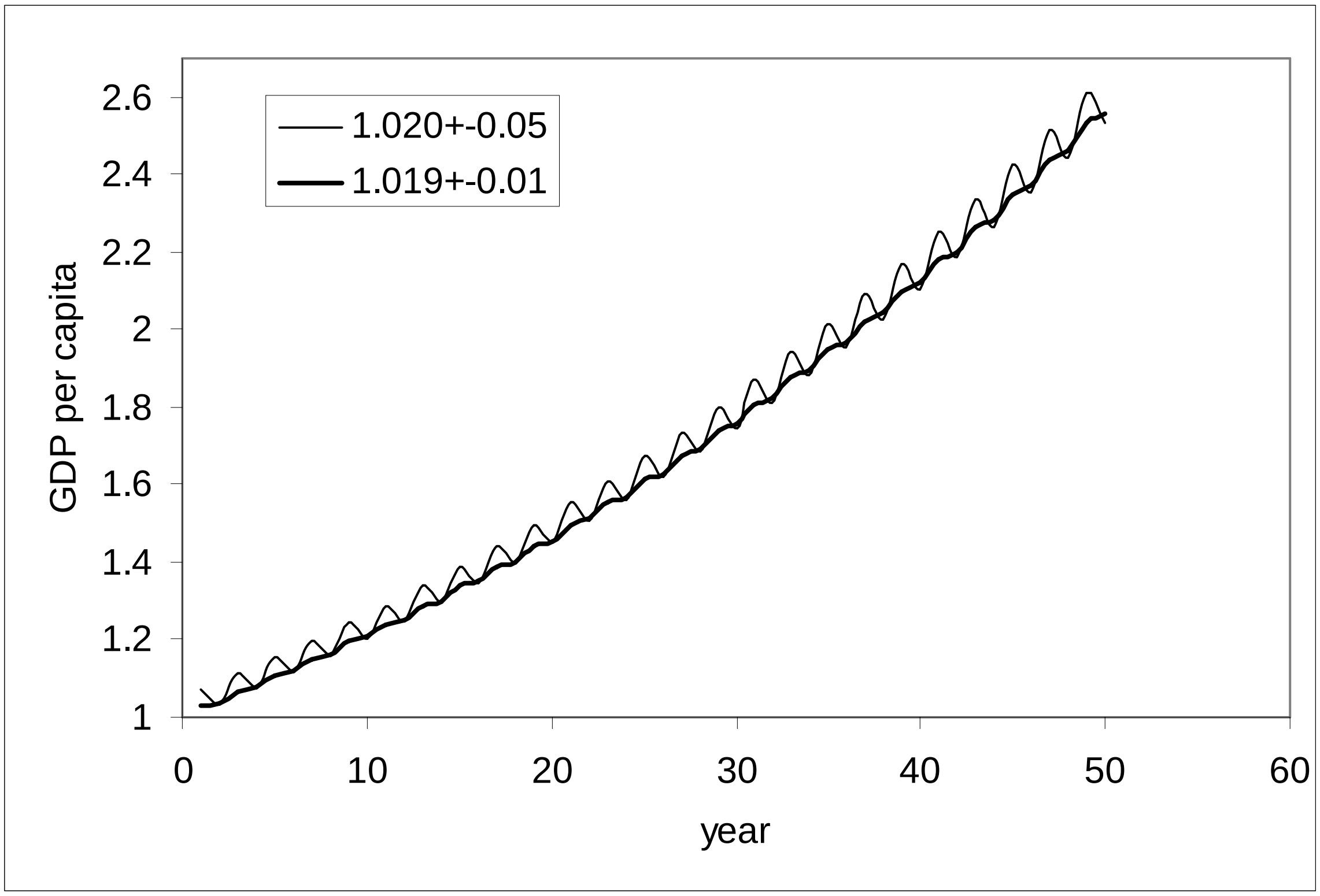

Fig. 29. Illustration of the effect of oscillation amplitude on growth rate. Two curves correspond to the mean growth value 1.02 with oscillations of ±0.05 (i.e. a sequence of growth rates 1.07, 0.97, 1.07 …) and the mean growth value 1.019 and oscillation amplitude of ±0.01. Being formally smaller ($1.019^{50}=2.56 < 2.69=1.02^{50}$), the latter case provides a larger total growth of 2.56 times compared to 2.53 in the former case. One must not average relative growth rates over long time intervals.

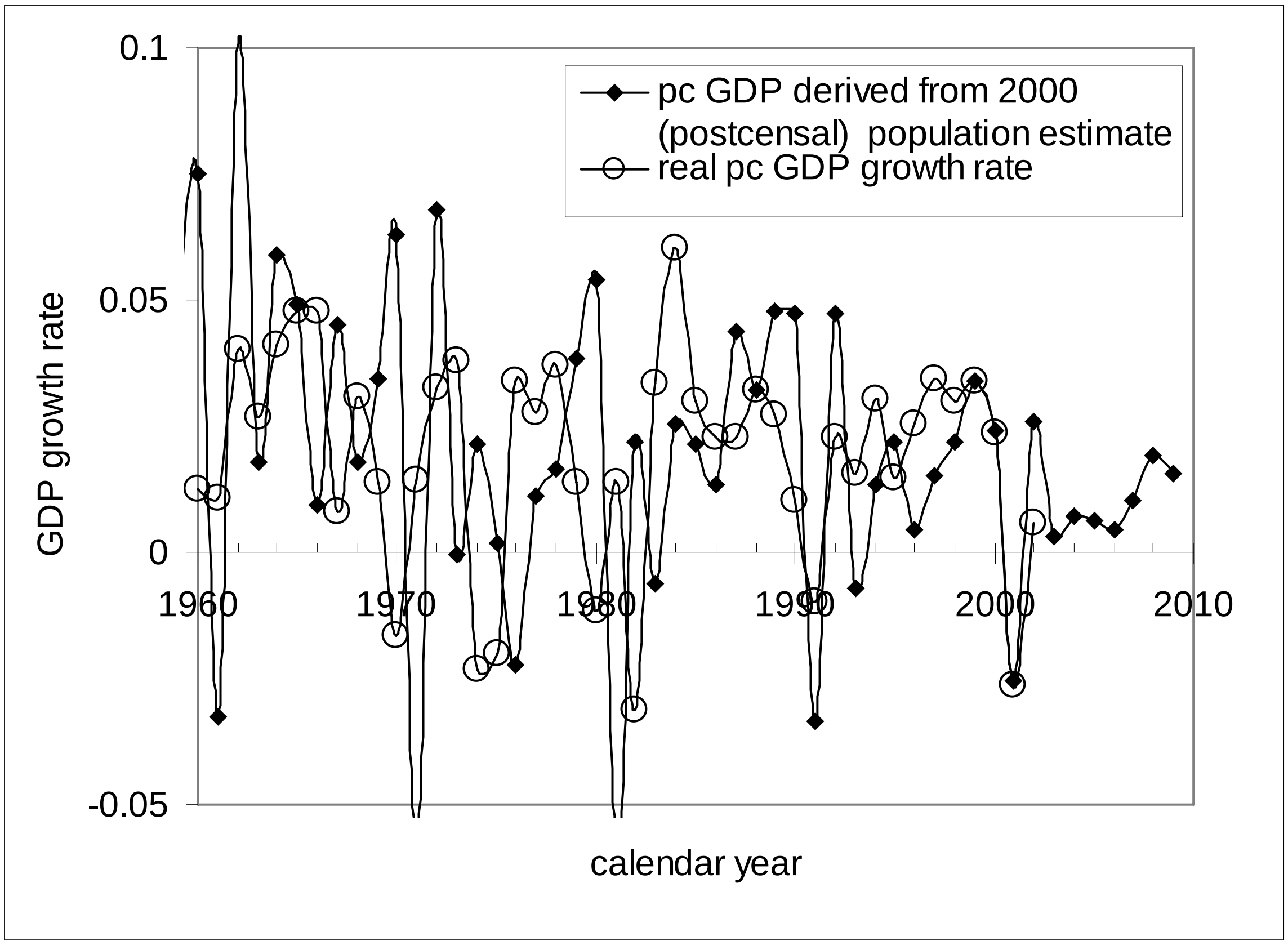

Fig. 30. Predicted by using relationship (7) and observed per capita GDP growth rates in the USA for the period between 1960 and 2002.

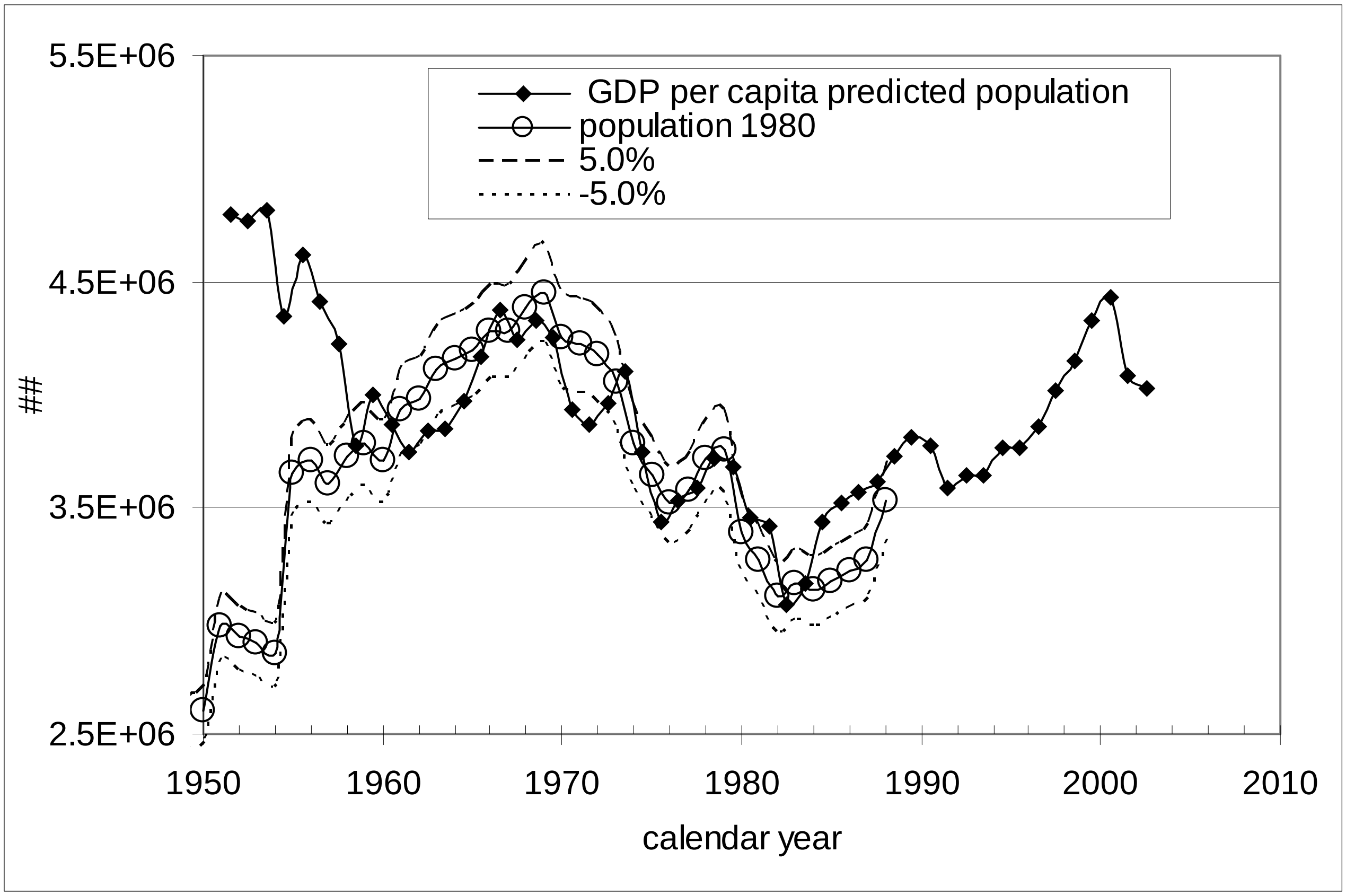

Fig. 31. Comparison of the 9-year-old population as obtained by a back projection of the 1980 population estimate and the number of 9-year-olds obtained from the per capita GDP observations according to equation (8).

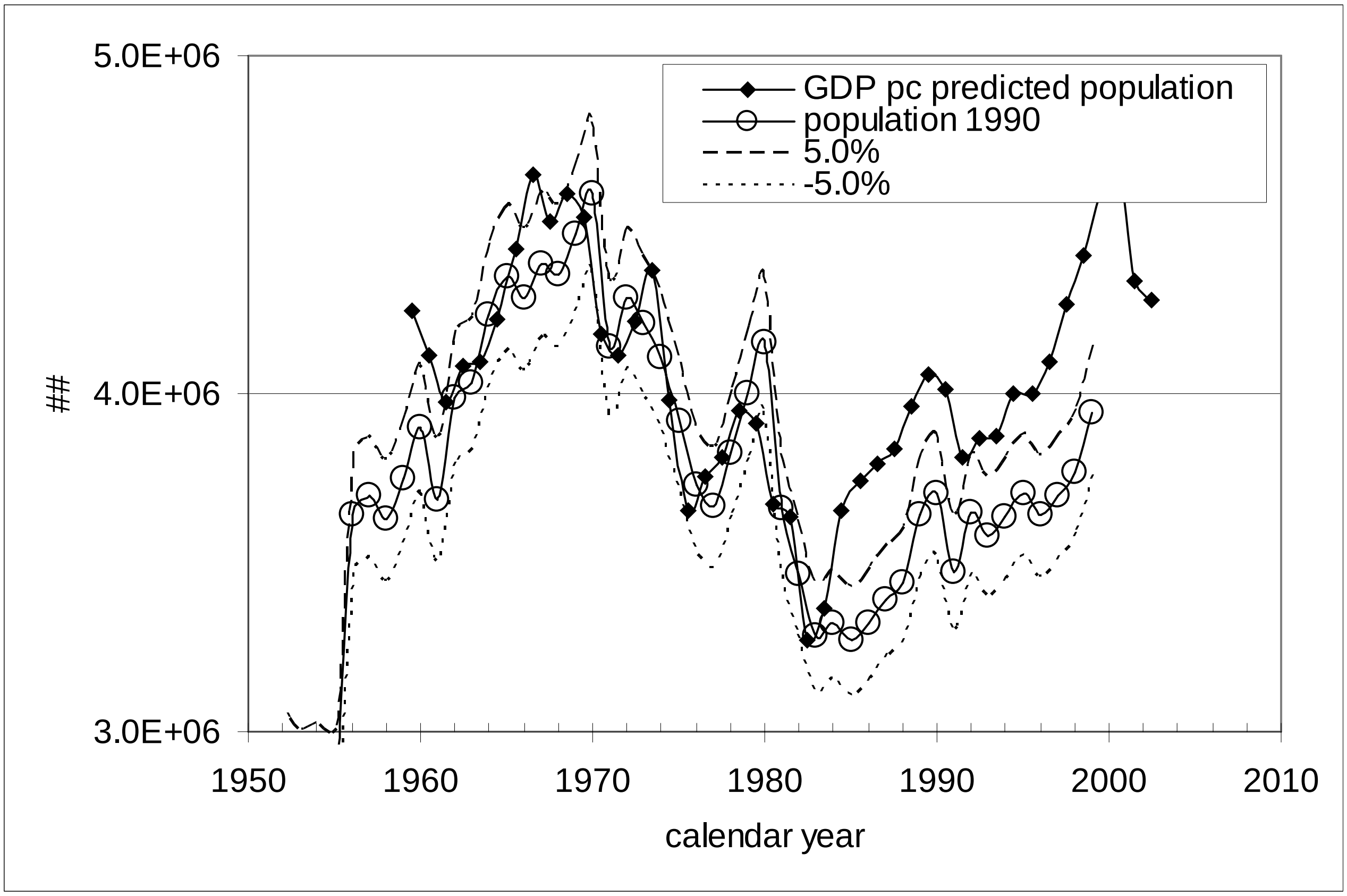

Fig. 32. Comparison of the 9-year-old population as obtained by a back projection f the 1990 population estimate and the number of 9-year-olds obtained from the per capita GDP observations according to equation (8).

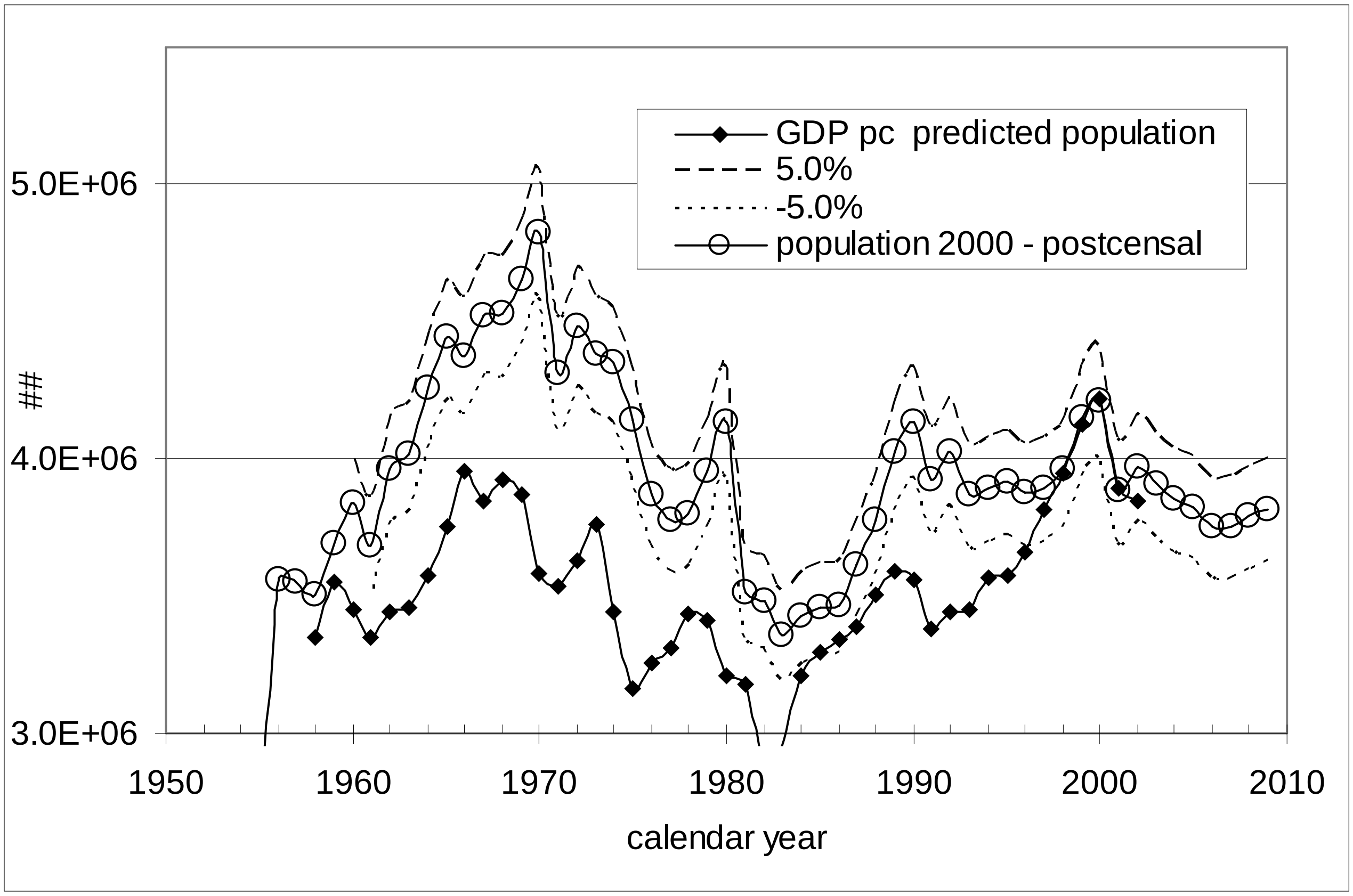

Fig. 33. Comparison of the 9-year-old population as obtained by a back projection of the 2000 postcensal population estimate and the number of 9-year-olds obtained from the per capita GDP observations according to equation (8).

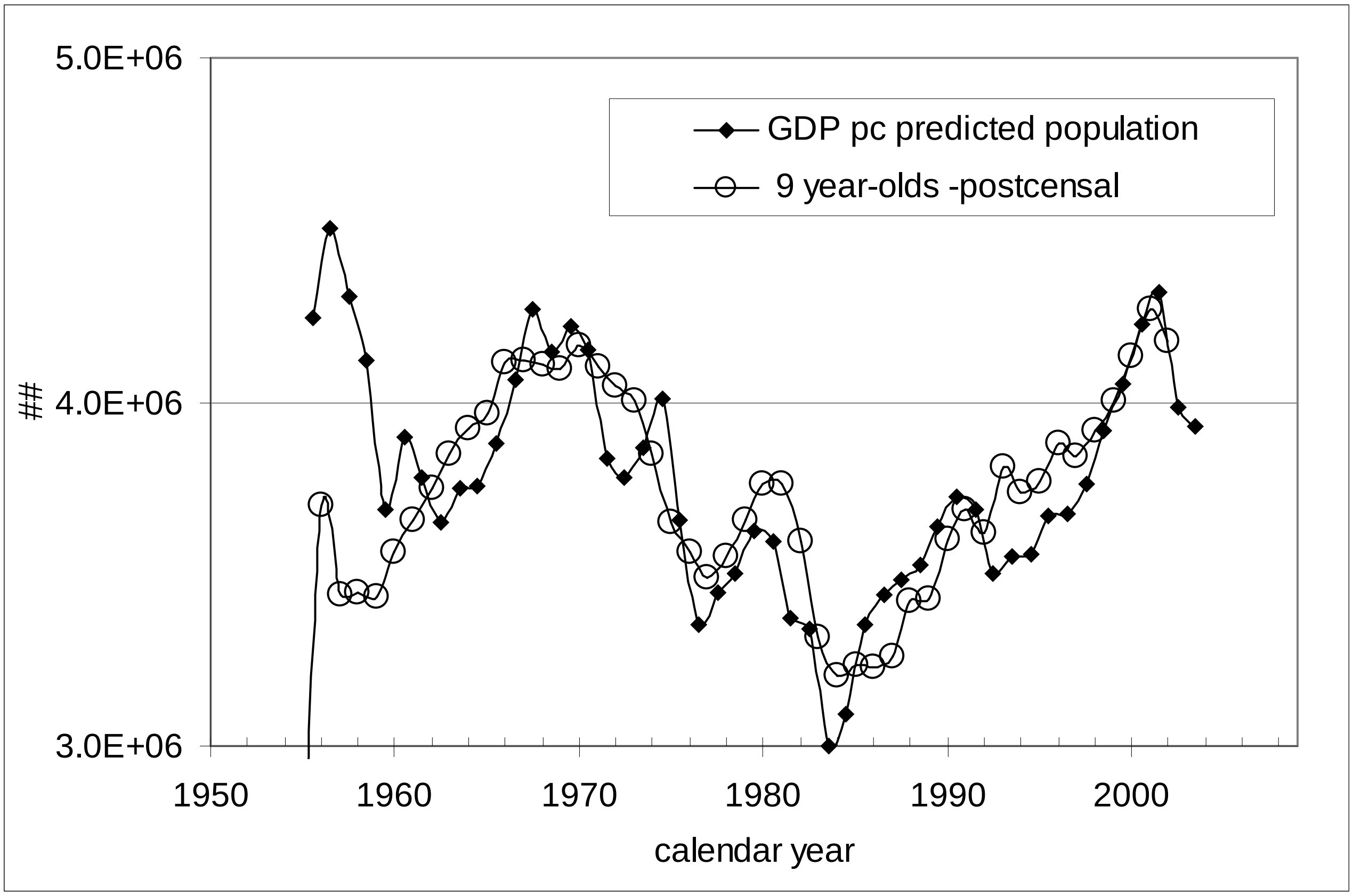

Fig. 34. Comparison of the 9-year-old population (postcensal estimate) and the number of 9-year-olds obtained from the per capita GDP observations according to equation (8). The observed behavior of the 9-year-old population is consistent with the per capita GDP observations.

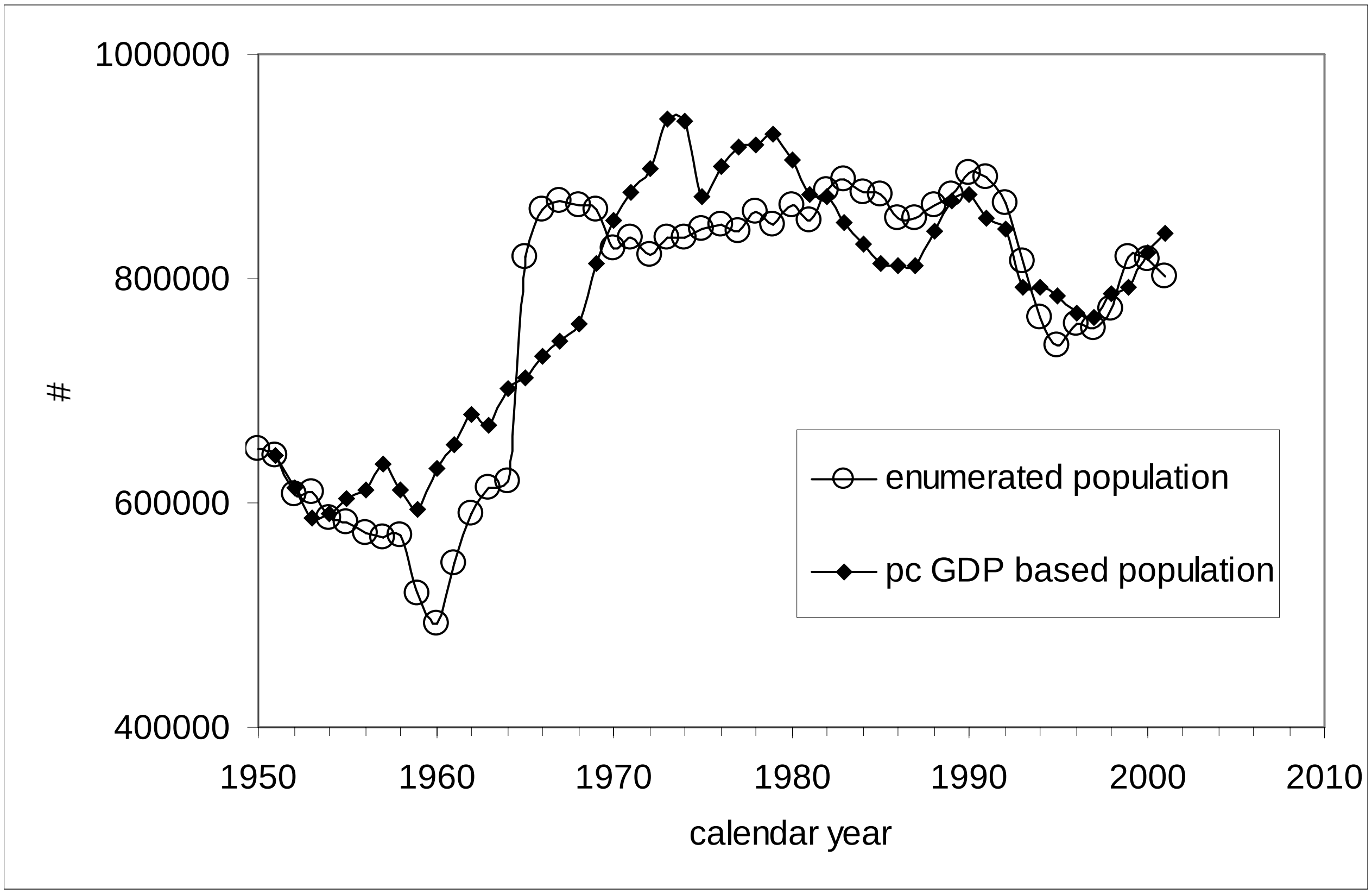

Fig. 35. Comparison of the 18-year-old population in France and the number of 18-year-olds obtained from the per capita GDP observations according to equation (8). The per capita GDP values are corrected for the difference between the total population and population of 15 year s of age and above.